\documentclass[a4paper,10pt]{article}
\usepackage[english]{babel}
\usepackage{url}
\usepackage{graphicx}
\usepackage{dcolumn}
\usepackage{bm}
\usepackage[]{amssymb,amsmath}
\usepackage{color}
\usepackage{cite}

\newcommand{\ER}{Erd\H{o}s\--R\' enyi}

\begin{document}

\title{Synergistic effects in threshold models on networks}

\author{Jonas S\o gaard Juul\thanks{Niels Bohr Institute, University of Copenhagen, Blegdamsvej 17, Copenhagen 2100-DK, Denmark}
\\ 
Mason A. Porter\thanks{Oxford Centre for Industrial and Applied Mathematics, Mathematical Institute, University of Oxford, Andrew Wiles Building, Radcliffe Observatory Quarter, Woodstock Road, Oxford, OX2 6GG, UK; CABDyN Complexity Centre, University of Oxford, Oxford, OX1 1HP, UK; and Department of Mathematics, University of California, Los Angeles, California 90095, USA}
}

\maketitle



\begin{abstract}

Network structure can have significant effects on the propagation of diseases, memes, and information on social networks. Such effects depend on the specific type of dynamical process that affects the nodes and edges of a network, and it is important to develop tractable models of spreading processes on networks to explore how network structure affects dynamics. In this paper, we incorporate the idea of \emph{synergy} into a two-state (``active'' or ``passive'') threshold model of social influence on networks. Our model's update rule is deterministic, and the influence of each meme-carrying (i.e., active) neighbor can --- depending on a parameter --- either be enhanced or inhibited by an amount that depends on the number of active neighbors of a node. Such a synergistic system models social behavior in which the willingness to adopt either accelerates or saturates depending on the number of neighbors who have adopted that behavior. We illustrate that the synergy parameter in our model has a crucial effect on system dynamics, as it determines whether degree-$k$ nodes are possible or impossible to activate. We simulate synergistic meme spreading on both random-graph models and networks constructed from empirical data. Using a local-tree approximation, we examine the spreading of synergistic memes and find good agreement on all but one of the networks on which we simulate spreading. We find for any network and for a broad family of synergistic models that one can predict which synergy-parameter values allow degree-$k$ nodes to be activated.

\end{abstract}



\vspace{1 cm}


\vspace{1 cm}



\begin{quotation}

{\bf Models of cascading processes on networks yield insights into a large variety of processes, ranging from the spread of information and memes in social networks to propagating failures in infrastructure and bank networks \cite{valente-book,rogers83,porter2016,rom-review2015,yamir-jcn2013,may2011}. In the context of social networks, there is long history of models of social influence based on overcoming individuals' thresholds with peer pressure or influence \cite{granovetter78,rogers83,valente-book,watts2002,kkt2003,dodds2005,centola2007}. Most such models consider peer pressure only from nearest neighbors, but it is also important to explore the influence of nodes beyond nearest neighbors (e.g., in the context of the ``three degrees of influence'' that has been reported in some studies) \cite{fowlerreview}. If the combined influence from several nodes is different than the sum of the influences from individual nodes, \textit{synergy} is taking place. Such synergistic effects can exert a major influence on spreading processes on networks. For example, in some systems, the amount of influence per person applying peer pressure may depend on the number of people who are applying peer pressure, and our goal in this paper is to incorporate such ideas into a threshold model of social influence in an analytically tractable way.  In our synergistic model, we examine social behavior in which the willingness to adopt either accelerates or saturates depending on the number of neighbors who have adopted some behavior. We illustrate that a synergy parameter can have a crucial effect on system dynamics (e.g., by determining whether degree-$k$ nodes are possible or impossible to activate). We also develop an analytical approximation that is effective at predicting the temporal development and cascade sizes in many networks.}

\end{quotation}


\section{Introduction}\label{intro}

Examining the spread of opinions, actions, memes, information, and misinformation in a population has received intense scrutiny in sociology, economics, computer science, physics, and many other fields \cite{porter2016,fowlerreview,yamir-jcn2013,granovetter78,valente-book,jackson2013,jackson2014,kkt2003,watts2002,loreto2009,Christakis07,centola2007,dodds2005,aral2009,ugander2012,goel-preprint,mollison1977}. Such phenomena --- including the spread of defaults of banks, norms in populations, and products or new practices in populations --- are often modeled as contagion processes that spread from node to node in a network \cite{elsinger_risk, Centola2005_norms, 10.1257/aer.99.5.1899}, in analogy with the spread of infectious diseases in a population. 

There are some similarities between social and biological contagions \cite{yy-epidemic}, and phrases like ``going viral'' arise from such similarities. Compartmental models, which were first developed in the context of biological contagions \cite{rom-review2015,porter2016}, are often used for modeling social contagions, although there are substantial differences between the spread of information and the spread of diseases \cite{Centola1194, PhysRevE.88.012818}. For example, media broadcasts affect the masses differently in social versus biological contagions \cite{goel-preprint}.

In addition to modeling spreading processes themselves, it is important to consider the effect of network structure on contagions \cite{porter2016,rom-review2015}. For example, it can have an important effect on phenomena such as the peak size and temporal development of outbreaks \cite{rom-review2015, taylor2015, colizza-prx2015, gleeson2013PRX, gleeson_2008_PRE, centola2007, gleeson-watts-weighted, MelnikChaos13, PhysRevE.88.012818, Centola2005_norms}. Various approaches have been used to understand such effects, including coupled differential equations, discrete dynamical systems, stochastic processes, agent-based models, and game theory \cite{porter2016,rom-review2015}. Of course, these different approaches are not completely independent from each other, as many have important connections to each other. For example, it is possible to construe a dynamical system on a network as an agent-based model (although the choice of different terminology often belies substantial differences in perspective) \cite{porter2016,loreto2009}, and some threshold models of contagions can also be derived from a game-theoretic perspective \cite{NWS:8888780}.

In the study of social contagions, many studies suppose that some small fraction of the nodes are infected initially, and they ask when a meme or disease can spread widely in a network \cite{porter2016,gleeson2013PRX}. When many nodes have adopted the meme (or become infected, in the context of a disease), it is said that a \textit{cascade} has occurred\cite{watts2002,goel-preprint}. A cascade can either be good or bad: a game developer may dream about his/her app becoming viral, but defaulting banks from systemic risk is a source of fear and dread in the financial sector. Seemingly viral spread of misinformation was also a prominent aspect of the 2016 U.S. presidential campaign and election. 

As our discussion suggests, in applications ranging from finance \cite{elsinger_risk} to meme spreading on Twitter\cite{PhysRevX.6.021019}, researchers are very interested in trying to identify what causes cascading behavior on networks \cite{goel-preprint}. In one prominent family of models, known as \emph{threshold models}, nodes survey their neighborhoods and adopt a meme (i.e., change their state) if sufficiently many of their neighboring nodes have already adopted this meme \cite{porter2016,granovetter78,valente-book, watts2002,gleeson2013PRX}. In most such models (and in most compartmental models), nodes are influenced only by their immediate neighbors, but in many situations (e.g., including social media such as Facebook and LinkedIn), individuals are able to observe actions by individuals beyond those to whom they are connected directly by an edge.\footnote{In fact, the sizes of the observable neighborhoods are different in different media (e.g., Facebook versus LinkedIn), and this can have profound effects on user experience, company algorithms, and more\cite{borgs2012}.}
In such situations, \textit{synergistic} effects can occur, as a node can be influenced by multiple nodes at the same time, and the combined influence differs from the sum of the individual influences. Synergistic effects can either increase or decrease the chance that a node will adopt a meme.

Synergistic effects can contribute to the dynamics of spreading processes in a diverse variety of contexts. Examples include the spread of behavior\cite{Centola1194}, the transmission of pathogens\cite{ludlam2012applications}, and the spread of new opportunities for farm activities among vineyards that form a wine route together\cite{brunori2000synergy}. Other phenomena with synergistic effects, which should be interesting to examine in the context of synergistic dynamical processes on networks, include the classical psychological ``sidewalk experiment'' with people staring up at the sky \cite{sidewalk}, increased value from the merging of companies (see, e.g., \cite{sudi1996}), and ``learning'' of delinquent and criminal behavior \cite{ballester2010}.

A few years ago, P\'erez-Reche et al. \cite{perez2011synergy} introduced a simple model of synergistic spreading in the context of a compartmental model for a biological contagion, and they examined its dynamics on a square lattice in two dimensions (2D). Their model was based on the standard susceptible--infectious--removed (SIR) model \cite{rom-review2015,porter2016}, in which an \textit{infectious} (I) node infects a \textit{susceptible} (S) neighbor at a constant rate $r_{\mathrm{SI}}=\alpha$. In this SIR model, an infectious node is infectious for a time $\tau$ before it switches states to \textit{removed} (R) (or ``recovered'', if one is less fatalistic), and then it can never become susceptible or infectious again. 

P\'erez-Reche et al. generalized this SIR model so that $r_{\mathrm{SI}}$ includes not only the parameter $\alpha$ but also a synergy term $r_{\mathrm{syn}}=\beta m_i$, where $m_i$ is the number of nodes that contribute to the synergy when updating node $i$.
They used a linear form of synergy: $r_{\mathrm{SI}}=\alpha+\beta m_i$. For $\beta<0$, the synergy is \textit{interfering}, as synergy lowers the chance that node $i$ becomes infectious; for $\beta>0$, the synergy is \textit{constructive}, as synergy increases the chance of node $i$ to become infectious. For $\beta=0$, the model in \cite{perez2011synergy} reduces to the standard SIR model, and there is no synergy. 

Several studies have followed up on the work of P\'erez-Reche et al. in \cite{perez2011synergy}. We mention some examples in passing now, and we give some more details in Section \ref{model-syn}. In \cite{Perez2013}, Taraskin et al. extended the theoretical analysis of \cite{perez2011synergy} and performed numerical computations in multiple types of $2D$ lattices. Reference \cite{PhysRevE.89.052811} studied a so-called ``generalized epidemic process'' (GEP), with interfering or constructive synergistic effects depending on the value of a parameter that models the amount of memory in social interactions. Reference~\cite{Perez2015} considered the effect of edge rewiring and nearest-neighbor synergy (so-called ``r-synergy'') on the invasiveness of diseases in various 2D lattices. Finally, a very recent paper \cite{liu2016explosive} explored a model for reversible synergistic spreading. The model was based on the susceptible--infectious--susceptible (SIS) model rather than the SIR model, and infectious nodes become susceptible again at some rate $\mu$. They defined synergistic effects (so-called ``d-synergy'') using a synergy parameter that depends on the next-nearest neighbors of a susceptible node. They found a critical value of their synergy parameter, above which the infectious fraction of nodes increases abruptly and dramatically. 

One thing that the above models have in common is that the update rules for node states include stochasticity. To facilitate analytical treatments of problems and to help isolate the effects of novel features in a model, it is often convenient to use deterministic update rules \cite{porter2016}. To better understand synergistic effects in spreading processes on networks, it is thus useful to examine such effects in models with deterministic update rules. By simplifying the framework in this way, we hope to improve understanding of synergistic effects in spreading processes. We will use a two-state deterministic model in the form of a linear threshold model \cite{granovetter78,valente-book,watts2002}, and in particular we will consider a binary (i.e., two-state) model in which a node can be \textit{active} or \textit{inactive}. In the context of social contagions, ``inactive'' nodes are susceptible, and ``active'' nodes are infected. Upon becoming infected, a node remains infected forever. We also focus on nearest-neighbor interactions (and, in particular, on what P\'erez-Reche call ``r-synergy'') although our approach is also amenable to models with next-nearest-neighbor interactions (what P\'erez-Reche call ``d-synergy'').

In the present paper, we introduce two models for the synergistic spread of memes on networks using threshold models with deterministic update rules. We develop analytical approximations for the spread of memes on networks constructed using a configuration model. To test our analytical approximations, we consider degree distributions from both empirical data and standard synthetic network models. We also compare the synergistic spread of memes on two empirical networks to configuration-model networks that we construct using degree distributions derived from the degree sequences of these two networks. We thereby hope to learn whether synergistic effects can produce different dynamics on empirical versus synthetic networks.

The rest of our paper is organized as follows. In Section \ref{model-syn}, we provide additional discussion of existing attempts to model synergy in spreading processes on networks. In Sections \ref{sec:synergistic}, \ref{sec:initial}, and \ref{sec:analytical}, we introduce our models for synergistic spreading on networks, examine this model on two empirical networks, and develop an analytical approximation to describe the fraction of activated nodes with degree $k$ and threshold $\phi$ in a network as a function of time. We also demonstrate that we expect certain values of a synergy parameter in the models to lead to abrupt changes in the dynamics. In Section \ref{synth}, we study synergistic spreading processes on several families of random networks. In Sections \ref{sec:3reg} and \ref{sec:ER}, we simulate synergistic spreading on $3$-regular and \ER{} (ER) random networks and compare our analytical approximation to the simulated spreading processes. In Section \ref{sec:realistic}, we simulate synergistic spreading on networks that we construct using the configuration model with degree distributions from two empirical networks. In all of these networks, we observe that our analytical approximation indicates when it becomes possible for a spreading meme to activate a node with degree $k$ and threshold $\phi$. We conclude in Section \ref{conc}.


\section{Modeling Synergy in Spreading Processes} \label{model-syn}

We now give some additional details about the model of synergistic spreading that was introduced by P\'erez-Reche et al. in \cite{perez2011synergy}. They defined two types of synergistic dynamics: (1) \textit{r-synergy}, in which $m_i+1$ is the total number of infectious nearest neighbors that simultaneously attempt to infect a focal susceptible node $i$; and (2) \textit{d-synergy}, in which $m_i$ is the number of infectious nodes that are connected to the infectious nearest-neighbor that attempts to infect the susceptible node $i$. 

In the simulations of P\'erez-Reche et al.~\cite{perez2011synergy}, only the node at the center of the square grid is infectious at time $t=0$; all other nodes start out in the susceptible state. P\'erez-Reche et al. called a disease ``invasive'' if it has a nonzero probability of reaching all four edges of the square grid before it is no longer possible to infect any other nodes. They illustrated that the value of a synergy parameter can affect whether a disease is invasive or noninvasive. They also illustrated that the value of a synergy parameter can affect whether an infectious host can infect more than one node. 
 
 Several papers have built on \cite{perez2011synergy} and produced additional insights on synergistic spreading dynamics on networks.  In \cite{Perez2013}, Taraskin et al. extended the theoretical analysis from \cite{perez2011synergy} by taking into account that the neighborhood of a node might change during its infectious period. They also simulated spreading via r-synergy on several types of 2D lattices. (Reference~\cite{perez2011synergy} considered only square lattices.) Each node in their lattices has the same degree (i.e., coordination number). They suggested that the synergy effects are most prominent in lattices with high node degree because of the increased number of possible contributors to the synergy effects. They also reported that lattices with high coordination number can have invasive synergistic diseases even when the transmission rate $\alpha \to 0$. 

Recently, reference\cite{Perez2015} considered the effect of r-synergy and rewiring of edges in various 2D lattices (square, triangular, and honeycomb) on the invasiveness of diseases. They examined a synergistic SIR model with three different expressions for the synergistic contribution to the infection rate. One of these expressions was the linear one introduced in \cite{perez2011synergy} and mentioned above, the other two were the exponential form $r_{SI} = \alpha e^{\beta n_i}$ and the corresponding linear approximation for small $\beta n_i$.

They considered spatial small-world (SSW) networks in which edges in a lattice are rewired from a neighbor to a ``nearby'' node (there is a maximum distance) with a certain probability and small-world (SW) networks in which there is no maximum distance for the rewiring. They studied the invasiveness of these synergistic contagions on these networks as a function of the number $k$ of nearest neighbors (i.e., coordination number) of the nodes in these different networks. In these networks, they reported that rewiring always lowers the rate $\alpha$ at which a contagion becomes invasive, independent of the value of the synergy parameter $\beta$ if the coordination number of the network is sufficiently small (e.g., $k=3$), that rewiring lowers the value of $\alpha$ at which contagions with interfering or low constructive synergy $\beta \in (0,\beta_*)$ become invasive regardless of the coordination number, and that rewiring increases the value of $\alpha$ at which contagions with a synergy parameter higher than a specific value (i.e., $\beta > \beta_* > 0$) become invasive in networks with sufficiently large coordination number (in particular, $k \ge 4$). 

Reference \cite{PhysRevE.89.052811} examined a so-called ``generalized epidemic process (GEP)'' with interfering or constructive synergistic effects depending on the value of a parameter that models the amount of memory in social interactions. In their GEP, the probability that a susceptible node is infected by an infectious neighbor depends on the number of previous unsuccessful attempts to infect that node. In their GEP, the first attempt to infect a node succeeds with a rate $\lambda$, and all subsequent attempts succeed with another rate $T$. Thus, the synergy is constructive for $T>\lambda$ but interfering for $T<\lambda$. Their updating rule differs slightly from those in the above studies: instead of infecting nodes that trying to infect their neighbors, susceptible nodes choose to ``adopt'' the state of a neighboring node with some probability. They interpreted constructive synergistic effects as social reinforcement, and they showed analytically that there is a continuous phase transition in the outbreak size for a family of modular networks when the social reinforcement is small. They constructed their modular networks by starting with $c$ complete subgraphs of equal size, and then rewiring each edge with independent probability $p$. Thus, lower values of $p$ correspond to more modular networks, and $p=1$ yields an \ER{} network.
Using the same family of networks, they also showed that their GEP undergoes a discontinuous phase transition in the contagion outbreak size when social reinforcement is high. 

Finally, a recent paper \cite{liu2016explosive} explored a variant of d-synergy. They used an SIS model to study the effect of d-synergy in networks. In contrast to the aforementioned studies, the spreading dynamics of this model is reversible, as infectious nodes eventually become susceptible again. They defined the probability that a susceptible node was successfully infected by an infectious node as $1-(1-\alpha)^{1+\beta n}$, where $n$ is the number of infectious nodes that are adjacent to the infectious node that is attempting to infect the susceptible node, $\alpha$ is the base transmission rate, and $\beta$ is the synergy parameter. As in standard SIS models, an infectious node becomes susceptible again at a rate $\mu$. They found a critical value for the synergy parameter in their model. Below this value, the steady-state density of infectious nodes increases continuously with the base infection rate. Above this parameter value, the steady-state density of infectious nodes in the network increases in an ``explosive'' manner (i.e., abruptly and drastically) as a function of their base infection rate. 

The models that we discussed above all have stochastic update rules, which can make it difficult to study models analytically. In the present paper, we consider synergetic dynamics in models with deterministic update rules. This facilitates analytical treatments, which we will use to shed light on synergistic spreading processes on networks.


\section{Synergistic Threshold Models}\label{sec:synergistic}

Perhaps the most popular type of deterministic model of meme spreading are \emph{threshold models} of social influence \cite{porter2016,granovetter78,valente-book,watts2002,kkt2003,centola2007}. In the simplest type of threshold model, which is a generalization of bootstrap percolation \cite{miller-roof2015,chalupa1979}, one chooses a threshold $\phi_i$ for each node independently from a probability distribution $f(\phi)$ at time $t = 0$ (in traditional bootstrap percolation, all nodes have the same threshold), and a node becomes ``active'' (i.e., it adopts the meme) if the fraction of its neighbors (or, in some variants, the number of its neighbors) that are active is at least this threshold. Because of the simplicity of basic threshold models, one can derive analytical approximations for cascade conditions in a variety of settings and in various extensions of the model \cite{watts2002,kkt2003,holme2005,gleeson2013PRX,MelnikChaos13}.

We seek to develop a synergistic threshold model. We focus on r-synergy and hence on nearest-neighbor interactions. (It is also worth thinking about d-synergy models, but we leave this for future work.) We examine networks that consist of unweighted, undirected $N$-node graphs. At each point in time, a node can be in one of two states: \textit{inactive} ($S_0$) or \textit{active} ($S_1$). Inactive nodes exert no influence on their neighbors, and active nodes exert some amount of influence on their neighbors. The total amount of influence exerted by all neighbors of a node $i$ gives the \emph{peer pressure} experienced by node $i$. Each node $i$ has a threshold $\phi_i$ drawn from a distribution $f(\phi)$ at time $t = 0$. We also activate a seed set of nodes at $t = 0$. In all of our simulations, the seed consists of a single node chosen uniformly at random. Whenever we consider updating node $i$ (which we do in discrete time with synchronous updating), it becomes active if and only if the peer pressure on it is at least $\phi_i$. 

We now construct a response function $F(n_i,k_i,\phi_i,\beta)$ that depends on the number $n_i$ of node $i$'s active neighbors, its degree $k_i$, its threshold $\phi_i$, and a synergy parameter $\beta$ that we will explain below. The response function is a non-decreasing function of $n_i$ and gives the probability that a node switches from the inactive state to the active one\cite{gleeson_2008_PRE}. One can use such a response function to describe numerous models of binary-state dynamics, such as bond and site percolation and the Watts threshold model (WTM) \cite{gleeson2013PRX}. We express the response function using a peer-pressure function $\Xi(n_i,\beta)$ by writing
\begin{equation} \label{eq:response_def}
	F(n_i,k_i,\phi_i,\beta) = 
	\begin{cases} 
0\,, & \text{if} \quad 	\Xi(n_i,\beta) < \phi_ik_i\,, \\
1\,, &  \text{otherwise}\,.
	\end{cases}
\end{equation}

We want to incorporate synergistic effects in $\Xi(n_i,\beta)$. P\'erez-Reche et al.~\cite{perez2011synergy} defined \textit{constructive synergy} and \textit{interfering synergy} by comparing their dynamics to a standard SIR model, which synergistic model generalizes. They defined the rate with which a susceptible node becomes infected as 
\begin{equation}
	\lambda = \max\{0, \alpha + (n_i-1) \beta\}\,,
\end{equation}
where $\alpha$ is a base infection rate in an SIR model without synergy, $\beta\in \mathbb{R}$ is a synergy parameter, and $n_i$ is the number of infectious nodes that exert synergistic influence on susceptible node $i$.

Whenever $\beta \neq 0$ and $n_i > 1$, this system exhibits synergy. If $\beta<0$, the synergy effects lower the rate with which susceptible nodes become infected from the combined effort of multiple infectious nodes exerting influence (as compared to the corresponding SIR model without synergy). Smaller (i.e., more negative) values of $\beta$ correspond to more powerful interfering synergy. In contrast, if $\beta>0$, the infection rate is larger, and larger $\beta$ results in more powerful constructive synergy. If $\beta = 0$, the infection rate is the same, and there is no synergy. Additionally, in P\'erez-Reche et al.'s model\cite{perez2011synergy}, synergy exists only if the number of nodes (called ``hosts'') that exert influence on a target node is strictly larger than $1$.
 If the number of hosts is $1$, the dynamics reduces to that of the corresponding standard SIR model. 

In the present paper, we draw inspiration from \cite{perez2011synergy} in terms of how we define interfering synergy and constructive synergy, but instead of generalizing a compartmental model of biological contagions, we start from the Watts threshold model (WTM) of social influence from \cite{watts2002}. 

In the WTM, $\Xi(n_i) = n_i$. We design two peer-pressure functions, which depend on the number $n_i$ of active neighbors and on a synergy parameter $\beta$. We require that
\begin{equation}
	\Xi(n_i,\beta) 
		\begin{cases}
			=0 \,, & \text{if } n_i= 0 \,, \\
			>n_i \,, & \text{if } \beta >0 \text{ and } n_i >1\,, \\
			= n_i \,, & \text{if } \beta = 0 \text{ or } n_i = 1\,, \\
			< n_i \,, & \text{if } \beta < 0 \text{ and } n_i > 1\,.
\end{cases}
\label{eq:model_demands}
\end{equation}
The two peer-pressure functions that we consider are
\begin{align} 
	\Xi_{\text{multiplicative}} &= (1+\beta)^{n_i-1}n_i \label{eq:multi_func} \,, \\ 
	\Xi_{\text{power}} &= n_i^{1+\beta} \label{eq:power_func} \,.
\end{align}
Naturally, these are not the only two functions that satisfy our demands in Eq.~\eqref{eq:model_demands}. In Section~\ref{sec:3reg}, we will argue that any non-synergistic peer-pressure function that is non-decreasing and continuous in the synergy parameter $\beta$ exhibits the same qualitative behavior as these two functions (in the sense of experiencing the same types of bifurcations).

If a node is \textit{vulnerable} (i.e., it can be activated by a single active neighbor), it remains vulnerable if one introduces synergy using Eq.~\eqref{eq:multi_func} or Eq.~\eqref{eq:power_func}. Moreover, no non-vulnerable node can become vulnerable as a result of the synergy introduced using Eq.~\eqref{eq:multi_func} or Eq.~\eqref{eq:power_func}. We seek to examine when synergy effects, as encapsulated by the parameter $\beta$, change the number of active neighbors that can activate a degree-$k$ node. That is, we seek to examine when synergy can assist or hinder the spread of a meme through a network. We can calculate when a specific change like this occurs. Suppose that a node $i$ with degree $k_i$ can be activated when there are at least $m_i$ active neighbors for $\beta = 0$. We wish to determine the $\beta$ values for which $l_i$ active neighbors are sufficient to activate node $i$. For the power synergy model \eqref{eq:power_func}, we calculate
\begin{align}
	(l_i)^{1+\beta} &\ge  \phi_ik_i\\
	\Rightarrow \beta &\ge \frac{\log \phi_i k_i}{\log(l_i)}-1\,. \label{eq:betacrit_power}
\end{align}
For multiplicative synergy model, we obtain 
\begin{equation}\label{eq:betacrit_multi}
	\beta \ge \left(\frac{\phi_i k_i}{l_i} \right)^{1/(l_i-1)}\,.
\end{equation}
More generally, except for $m_i=1$ or $l_i = 1$ (by construction, nodes cannot become or stop being vulnerable from synergistic effects), we can solve for the value at which any $l_i \in \mathbb{N}$ active neighbors can activate a node with degree $k_i$ and threshold $\phi_i$, given the synergy parameter $\beta$. Hence, the threshold $\phi_i$ exclusively determines if a node is vulnerable. If a node is not vulnerable, the synergy parameter can alter the difficulty with which it is activated for any threshold $\phi_i$. 

When we initiate our simulations with only a single node as a seed, there is a risk that this seed is surrounded --- or is part of a group of vulnerable nodes of insignificant size that are surrounded --- by non-vulnerable nodes. Because such situations arise from the choice of threshold distribution $f(\phi)$ rather than from synergistic effects, we discard such simulations throughout this paper.


\section{Synergy in two Empirical Networks}\label{sec:initial}

We start by examining the synergistic threshold model \eqref{eq:power_func} on the network of condensed-matter physics paper coauthorships from \cite{leskovec_graph_2007}. (The network is available at \url{https://snap.stanford.edu/data/}.) In this network, a node represents an author, and there is an (undirected) edge between nodes $i$ and $j$ if the authors coauthored at least $1$ paper. We suppose for simplicity that all nodes have a threshold of $\phi^* = 1/10$. 

\begin{figure*}[tb]
\includegraphics[width=.49\linewidth]{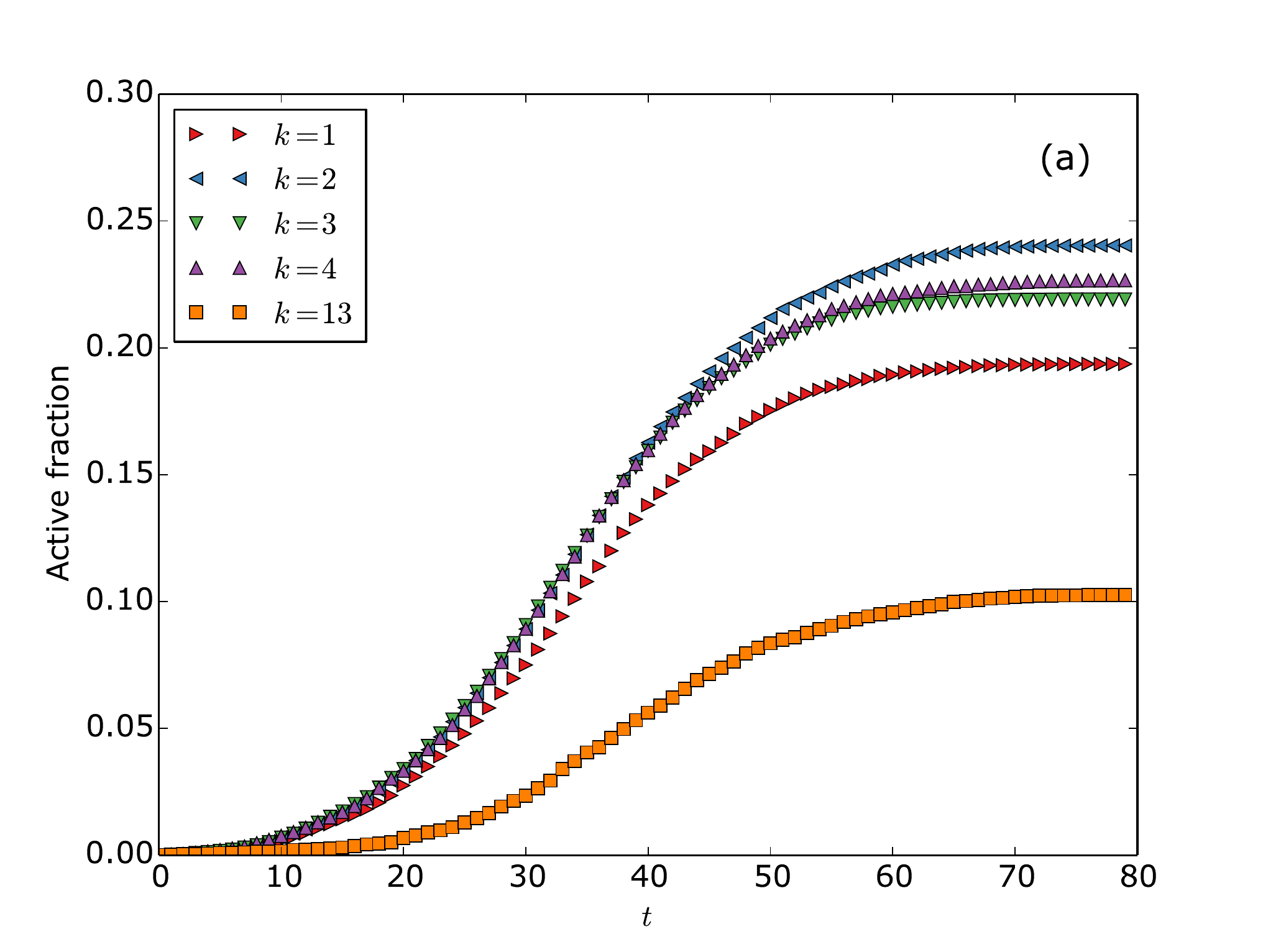}
\hfill
\includegraphics[width=.49\linewidth]{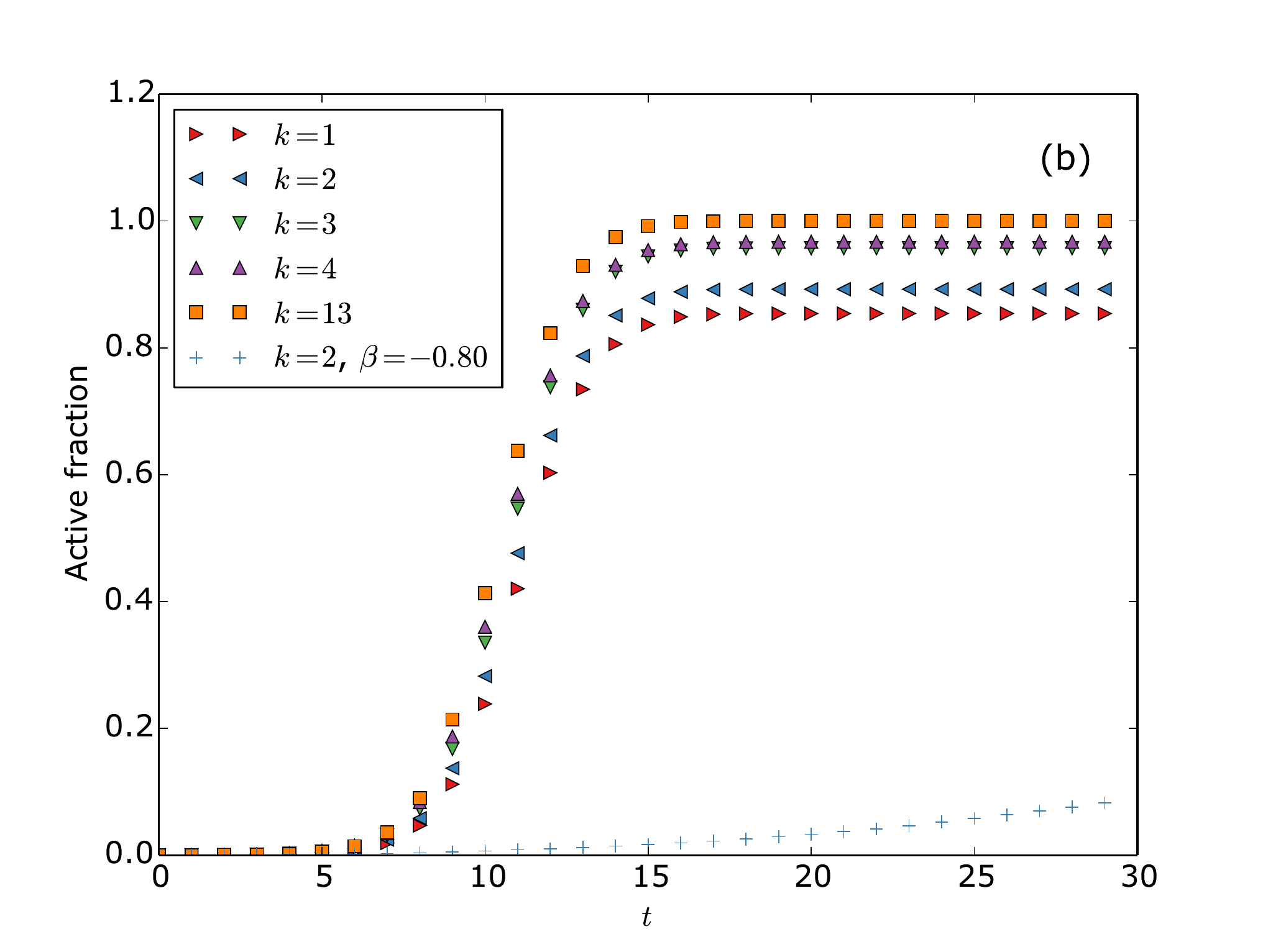}
\caption{Example of the behavior of the synergistic threshold model defined with \eqref{eq:power_func} using (a) interfering synergy (with $\beta = -0.80$) and (b) constructive synergy (with $\beta = 0.15$). In panel (b), we show part of the curve for $k=2$ from the case of interfering synergy for comparison.  Because we choose the seed active node uniformly at random, there is a chance that only the seed is activated. We do not take such runs into consideration. For the interfering synergy plot, only the seed was activated in $94$ of $110$ runs; for constructive synergy, this occurred in $31$ of $110$ runs. For the simulations in this figure, we ran the synergistic threshold model on the condensed-matter physics coauthor network from \cite{leskovec_graph_2007}, and the threshold for each node is $\phi = \phi^* = 1/10$. For each degree, a smaller fraction of nodes become active for interfering synergy than for constructive synergu. We also see that it takes longer for the meme to spread in the network for interfering synergy than for constructive synergy. 
}
\label{fig:condmat_actual}
\end{figure*}

We show our results in Fig.~\ref{fig:condmat_actual}. We use power synergy \eqref{eq:power_func}, and we show interfering synergy ($\beta = -0.80$) in the left panel (a) and constructive synergy ($\beta = 0.15$) in the right panel (b). Data points correspond to the mean fraction of degree $k$ nodes that are active at each time step in question. Among our simulations, we include only realizations in which the meme activates at least $0.5\%$ of the network. For each degree, an equally large or smaller fraction of nodes is activated for interfering synergy than for constructive synergy. In panel (b), we show the $k=2$ curve from panel (a) for comparison. We see that it takes longer for the meme to spread in the network for interfering synergy than it does for constructive synergy.

\begin{figure*}[tb]
\includegraphics[width=0.49\linewidth]{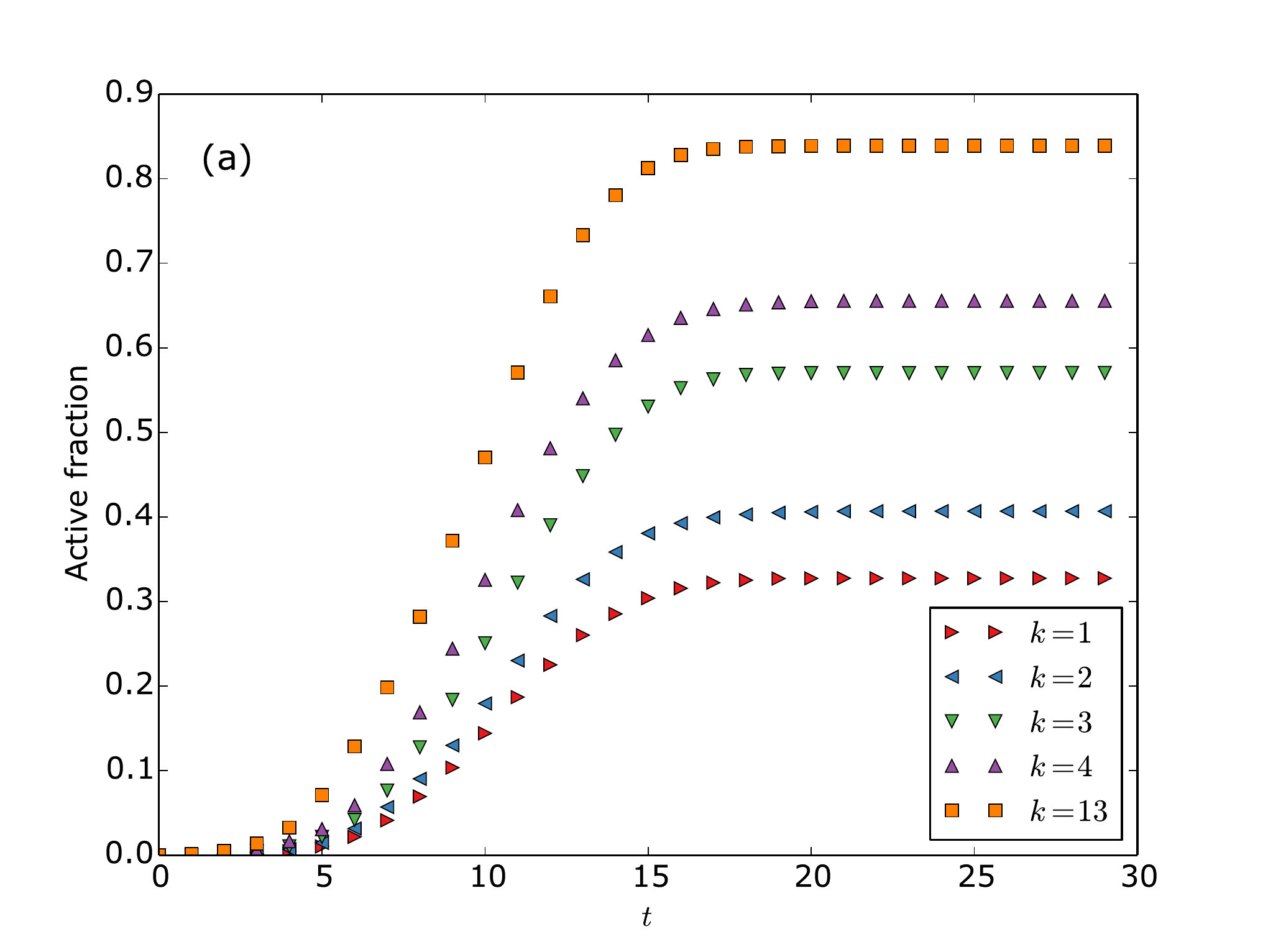}
\hfill
\includegraphics[width=0.49\linewidth]{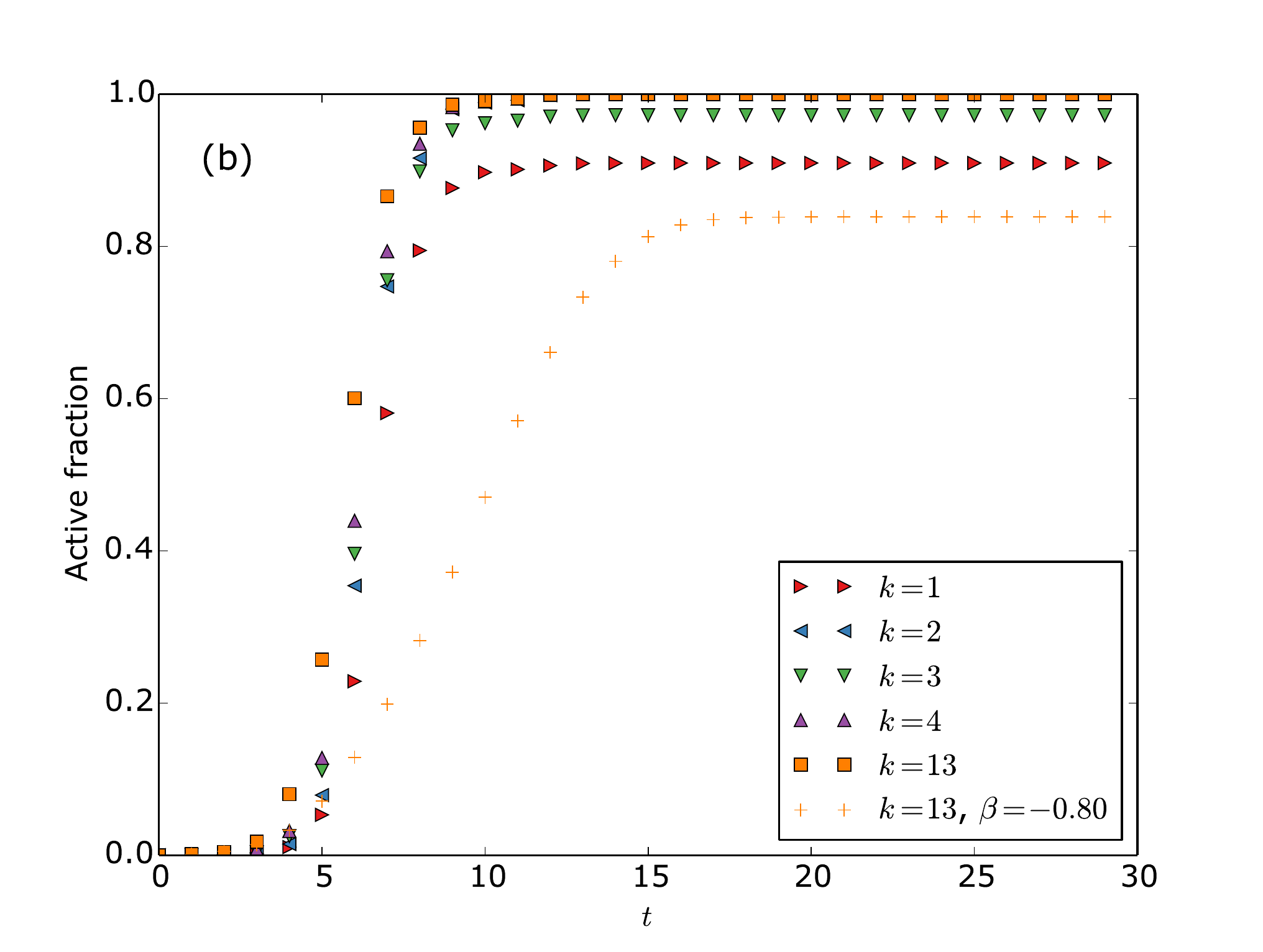}
\caption{Example of the behavior of the synergistic threshold model defined with \eqref{eq:power_func} using (a) interfering synergy (with $\beta = -0.80$) and (b) constructive synergy (with $\beta = 0.15$). In panel (b), we show the curve for $k=13$ for the case of interfering synergy for comparison.  Because we choose the seed active node uniformly at random, there is a chance that only the seed is activated. We do not take such runs into consideration. For the interfering synergy plot, only the seed was activated in $30$ of $110$ runs; for constructive synergy, this occurred in $24$ of $110$ runs. In the simulations in this figure, we ran the synergistic threshold model on the {\sc Northwestern25} network from the {\sc Facebook100} data set \cite{traud_social_2012}, and the threshold of each node is  $\phi = \phi^* = 1/33$. For each degree, a smaller fraction of nodes becomes active for interfering synergy than for constructive synergy. We also see that it takes longer for the meme to spread in the network for interfering synergy than it does for constructive synergy. 
}\label{fig:facebook_original}
\end{figure*}

We now examine our synergistic threshold model on another empirical network, the {\sc Northwestern25} network from the {\sc Facebook100} data set\cite{traud_social_2012}. This data set contains the complete set of people and friendships of $100$ different U.S. universities from one day in autumn $2005$. {\sc Northwestern25} is the data from Northwestern University. We show our results in Fig.~\ref{fig:facebook_original}. We suppose that all nodes have a threshold of $\phi^* = 1/33$, and we again examine power synergy with interfering synergy (with $\beta = -0.80$) in panel (a) and constructive synergy (with $\beta = 0.15$) in panel (b). For comparison, we include the curve for $k=13$ for constructive synergy for our plots for interfering synergy. We again see that it takes longer for the meme to spread in the network for interfering synergy than it does for constructive synergy, and that the fraction of nodes that are active is smaller or of equal size for interfering synergy than it is for constructive synergy.

\section{Analytical Approximation of Number of Active Nodes Versus Time}\label{sec:analytical}

We now develop an analytical approximation that describes the fraction of active nodes in a network as a function of time for any choice of peer-pressure function, degree distribution, and threshold distribution.

Recall that we employ synchronous updating in our simulations. Because our model is deterministic, this choice does not affect the final infected fraction of active nodes. We activate $1$ seed node of the $N$ total nodes at time $t=0$, and it is convenient for the theory to express it as a fraction $\psi_k^{\phi} = 1/N$ of the nodes with degree $k$ and threshold $\phi$. See \cite{gleeson_2007_PRE, gleeson_2008_PRE} for a discussion of the effects on cascade side of using a single active node as a seed for the WTM, and see \cite{fennell2016} for a recent discussion of issues with synchronous versus asynchronous updating (where asynchronous updating, such as through a Gillespie algorithm, is meant to model continuous-time dynamics) for dynamical processes on networks.

To calculate the fraction $\rho_k^\phi$(n+1) of active nodes with degree $k$ and threshold $\phi$ at time step $n+1$, we write the recursive formula (as in, e.g., \cite{gleeson_2007_PRE,gleeson_2008_PRE,MelnikChaos13})
\begin{equation}\label{eq:rhok}
	\rho_k^{(\phi)}(n+1) = \psi_k^{(\phi)} + (1-\psi_k^{(\phi)})\sum_{j=0}^{k}B^k_j(\bar{q_k}^{(\phi)}(n))F(j_i,k_i,\phi_i,\beta)\,,
\end{equation}
where $\bar{q}_k^{(\phi)}(n)$ is the probability that a neighbor of an inactive node with degree $k$ and threshold $\phi$ chosen uniformly at random is active at time step $n$, and
\begin{equation}
	B^k_j(p) = \binom{k}{j}p^j(1-p)^{k-j}\,.
\end{equation}

We can write $\bar{q}_k^{(\phi)}(n)$ as a function of $q_{k'}^{(\phi')}(n)$, the probability that, for a given inactive node, a neighbor with degree $k'$ and threshold $\phi'$ is active at time step $n$. This probability is
\begin{equation}
	\bar{q}_k^{(\phi)}(n) = \frac{\sum_{k',\phi'}P\left((k,\phi),(k',\phi')\right)q_{k'}^{\phi'}(n)}{\sum_{k',\phi'}P\left((k,\phi),(k',\phi')\right)}\,,
\end{equation}
where $P\left((k,\phi),(k',\phi')\right)$ is the probability that a node with degree $k$ and threshold $\phi$ is adjacent to a node with degree $k'$ and threshold $\phi'$. For an inactive node, the probability that a neighboring node with degree $k$ and threshold $\phi$ is active is
\begin{equation}\label{eq:qk}
	q_k^{(\phi)}(n+1) = \psi_k^{(\phi)}+(1-\psi_k^{(\phi)})\sum_{j=0}^{k-1}B^k_j(\bar{q_k}^{(\phi)}(n))F(j_i,k_i,\phi_i,\beta)\,.
\end{equation}
The only difference between Eq.~\eqref{eq:qk} and Eq.~\eqref{eq:rhok} stems from the fact that the degree-$k$ neighbor, which we consider in \eqref{eq:qk}, has a maximum of $k-1$ active neighbors if it is adjacent to at least one inactive node. In these equations, we have assumed that each neighbor of node $i$ is independent of the others, so we are assuming that the network is locally tree-like \cite{localtreeapprox_mason,porter2016}


\section{Synergy in Synthetic Networks}\label{synth}

To illustrate our theoretical results, we consider synergistic spreading in several families of random graphs.


\subsection{Synergy in 3-Regular Networks}\label{sec:3reg}

We first examine 3-regular random networks, in which every node has degree 3 and stubs (i.e., ends of edges) are matched uniformly at random. That is, we consider configuration-model networks in which each node has degree 3. We study how synergy effects influence the spread of memes on these networks by examining several values of the parameter $\beta$ for both multiplicative and power synergy. In our numerical simulations, we suppose that a fraction $p_0 = 0.8$ of the nodes have threshold $\phi  = 0.32 < 1/3$ and that a fraction $1 - p_0 = 0.2$ of the nodes have threshold $\phi =1$. 

\begin{figure}[tb]
\centering
\includegraphics[width=\linewidth]{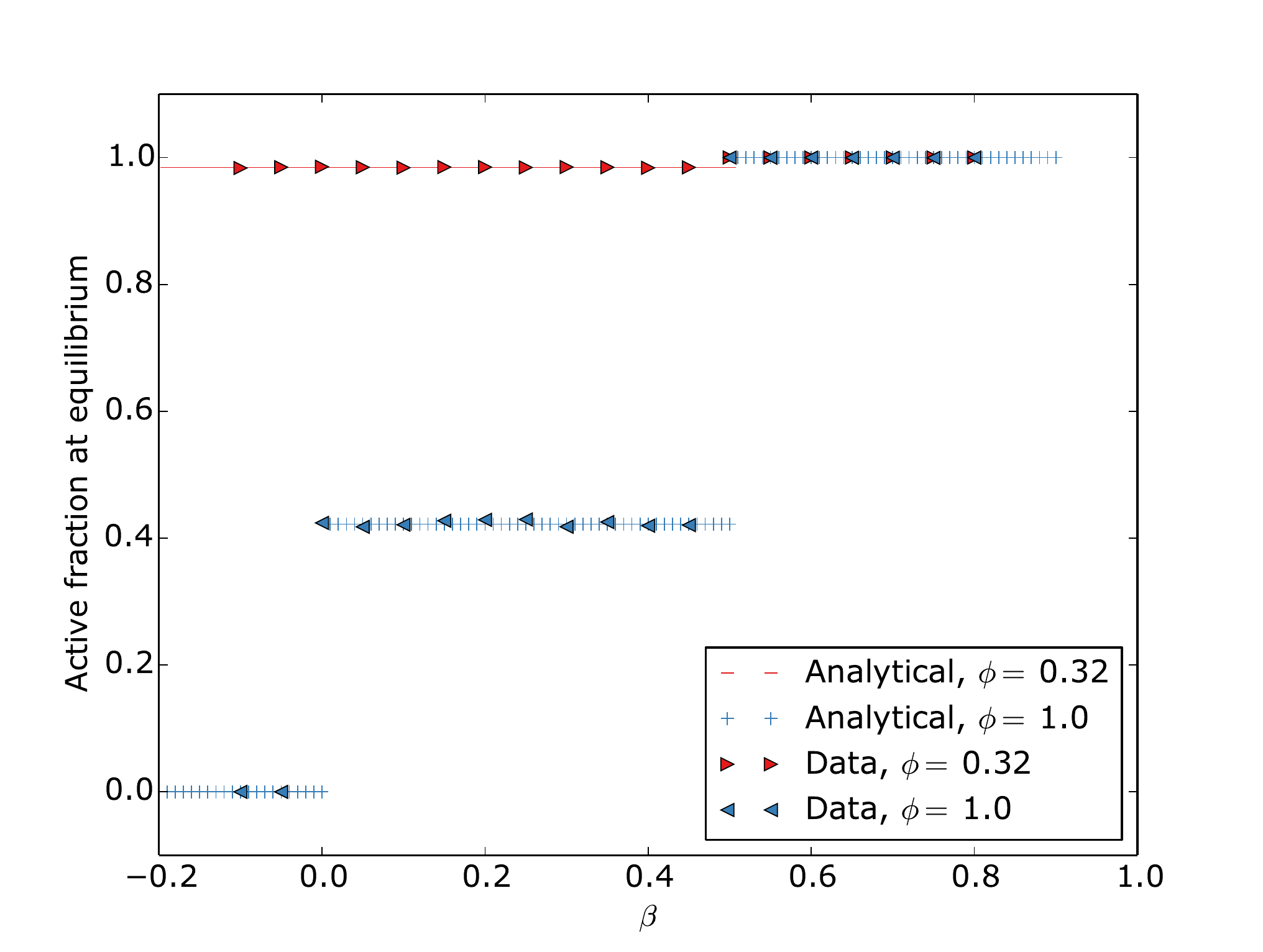}
\caption{Final fraction of active nodes in $3$-regular random networks of $10,000$ nodes when using the multiplicative synergistic peer-pressure function \eqref{eq:multi_func}. A fraction $p_0=0.8$ of the nodes have threshold $\phi_0  = 0.32 < \frac{1}{3}$, and a fraction $1-p_0 = 0.2$ of the nodes have threshold $\phi = 1$. Each data point is a mean of $10$ realizations of the synergistic threshold model on $10$ different $3$-regular random networks, which we create using a configuration model. For each $\beta$ value, we create $10$ networks. (In doing these simulations, we discarded $2$ realizations due to the choice of seed node; the contagion did not spread enough in those cases.)
}
\label{fig:t_inf}
\end{figure}

In all networks from this point onwards, we create a new network for each realization of a synergistic threshold model. For all networks except \ER (ER) networks, we specify a degree distribution $p(k)$. We use this to determine a degree for each of $10,000$ degrees, and we then connect these nodes to each other using a configuration model (connecting stubs to each other uniformly at random) \cite{Fosdick2016}.
We choose a single node uniformly at random as a seed and update nodes synchronously at each discrete time step. We stop the simulations only when we reach equilibrium (i.e., when no more nodes can eventually activate). In Fig.~\ref{fig:t_inf}, we plot the equilibrium active fractions of high-threshold and low-threshold nodes as a function of the synergy parameter $\beta$. Each data point is a mean over $10$ realizations of the spreading process.

When $\beta$ surpasses the values $0$ and $0.5$, the final fraction of active nodes with threshold $\phi = 1$ increases dramatically. We can see this from Eqs.~\eqref{eq:betacrit_multi} and \eqref{eq:response_def}. For $\beta <0$, it is not possible to satisfy $\phi_i k_i \ge (1+\beta)^{n_i-1}n_i$, because $n_i\le k_i$. For $\beta \in [0,0.5)$, the relation $\phi_i k_i \ge (1+\beta)^{n_i-1}n_i$ holds only for $n_i=k_i$. 
In this case, nodes with $\phi=1$ can be activated, but they are never able to help activate a neighbor (unless they are part of the seed set of active nodes), as all of their neighbors are necessarily already active once they have been activated. For $\beta \ge 0.5$, the relation $\phi_i k_i \ge (1+\beta)^{n_i-1}n_i $ holds for $n_i = k_i$ and $n_i=k_i-1$. In this case, nodes with $\phi=1$ can be activated even when they still have an inactive neighbor. Hence, nodes with $\phi=1$ can help spread the meme, resulting in an increase in active nodes with both $\phi=1$ and $\phi = 0.32$ compared to what occurs for $\beta < 0.5$. Rephrasing these observations, bifurcations occur at special values of $\beta$ (which are $\beta = 0$ and $\beta = 0.5$ in this example) for the peer-pressure function \eqref{eq:multi_func}, and we calculate the bifurcation points by solving $\Xi(n_i) = k_i\phi_i$ for $n_i \in \{2,\ldots, k_i \}$ (where we exclude $n_i = 1$ because it corresponds to a vulnerable node, which by design, would be vulnerable for any value of $\beta$). Such $\beta$ values exist for any non-decreasing peer-pressure function $\Xi(n_i,\beta)$ that is continuous in $\beta$. For two different peer-pressure functions, the $\beta$ value that makes it possible for a specific number (e.g., 4, to be concrete) active neighbors to activate a specific node can differ, but there is some value of $\beta$ in both peer-pressure functions. Hence, all continuous, non-decreasing synergistic peer-pressure functions behave in qualitatively the same way. 

\begin{figure}[tb]
\centering
\includegraphics[width=\linewidth]{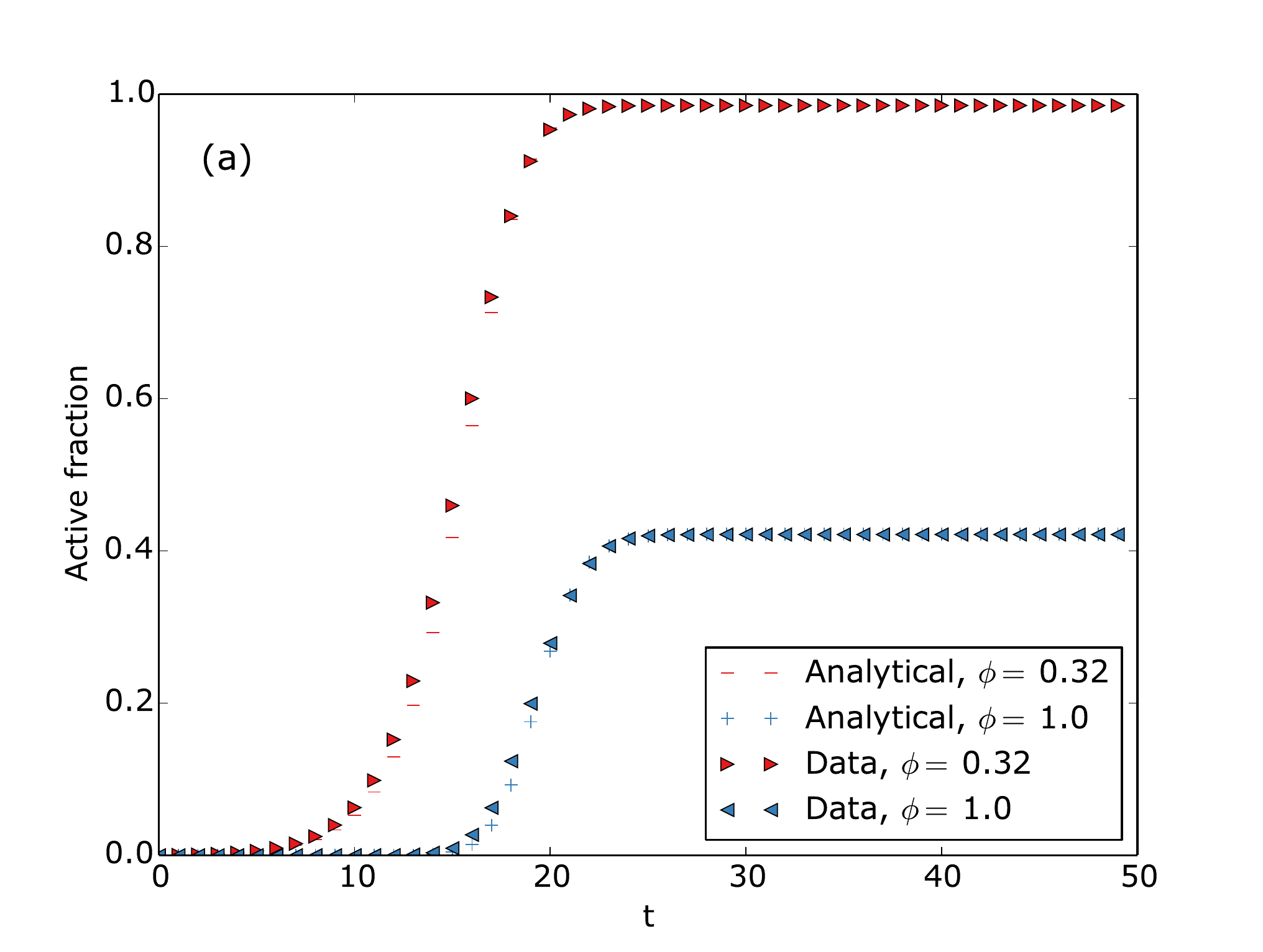}
\includegraphics[width=\linewidth]{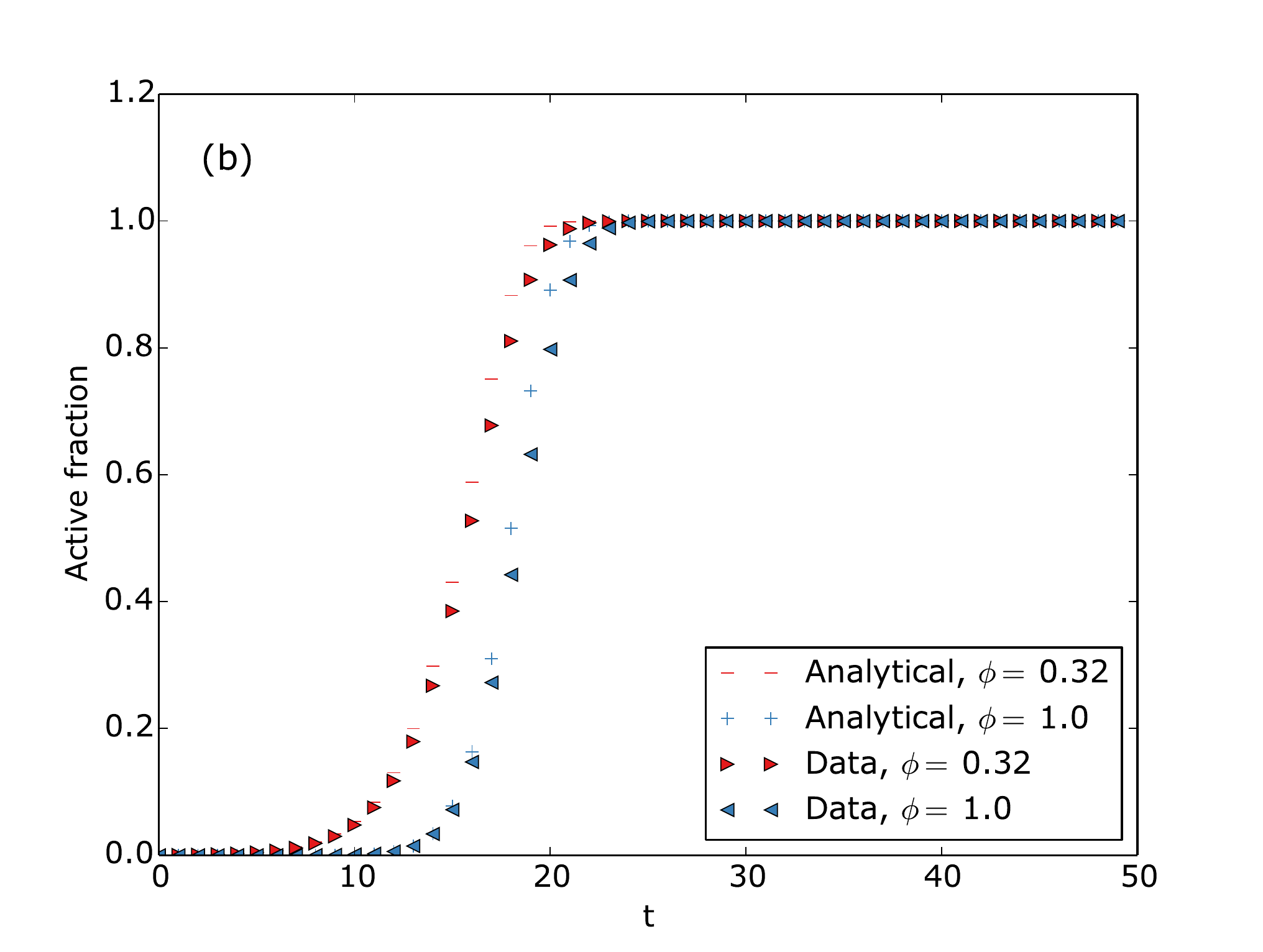}
\caption{Active fraction of nodes as a function of time in $3$-regular random networks of $10,000$ nodes. A fraction $p_0=0.8$ of the nodes have threshold $\phi_0  = 0.32 < \frac{1}{3}$, and a fraction $1-p_0 = 0.2$ have threshold $\psi=1$. In panel (a), the synergy parameter is $\beta = 0.4999$; in panel (b), it is $\beta = 0.5001$. In each panel, each data point is a mean over $100$ realizations of memes that spread using the synergistic response function with peer-pressure function \eqref{eq:multi_func}. We observe excellent agreement between the analytical approximation \eqref{eq:rhok} and our simulations. (In these simulations, we did not need to discard any realizations due to the choice of seed node.) For each realization, we created a $3$-regular random network using a configuration model. The two panels show results for two different sets of networks.
}
\label{fig:3reg_small}
\end{figure}

In Figs.~\ref{fig:3reg_small}(a) and \ref{fig:3reg_small}(b), we show how the meme spreads for $\beta = 0.4999$ and $\beta = 0.5001$, respectively. Each data point is a mean over $100$ realizations of the spreading process. For each realization, we create a new $3$-regular random network using a configuration model (with stubs connected uniformly at random).

One can use any response function, such as ones that use the peer-pressure functions \eqref{eq:multi_func} or \eqref{eq:power_func}, to compute when $n_i\le k_i$ nodes can activate a node with threshold $\phi_i$ by solving the equation $\Xi(n_i) = \phi_ik_i$. Therefore, different response functions can have sudden increases in the final fraction of active nodes at critical values of $\beta$ for the same reason: at these values of $\beta$, it becomes possible for some nodes to be activated with fewer active neighbors than was the case for smaller values of $\beta$. Although these critical values of $\beta$ can differ for different response functions, the different synergistic response functions exhibit qualitatively similar behavior. Therefore, we henceforth use only the response function that is specified by the power peer-pressure function \eqref{eq:power_func}.

\subsection{Synergy in \ER {} Networks}\label{sec:ER}

We now simulate the spread of memes with synergy on ER networks. First, we consider ER networks with mean degree $z=3$, and we then consider ER networks with mean degree $z=8$. In both cases, we use the threshold distribution $f(\phi) = \delta(\phi-\phi^*)$ with $\phi^* = 1/7$.


\subsubsection{Mean Degree $z=3$}

\begin{figure}[tb]
\centering
\includegraphics[width=\linewidth]{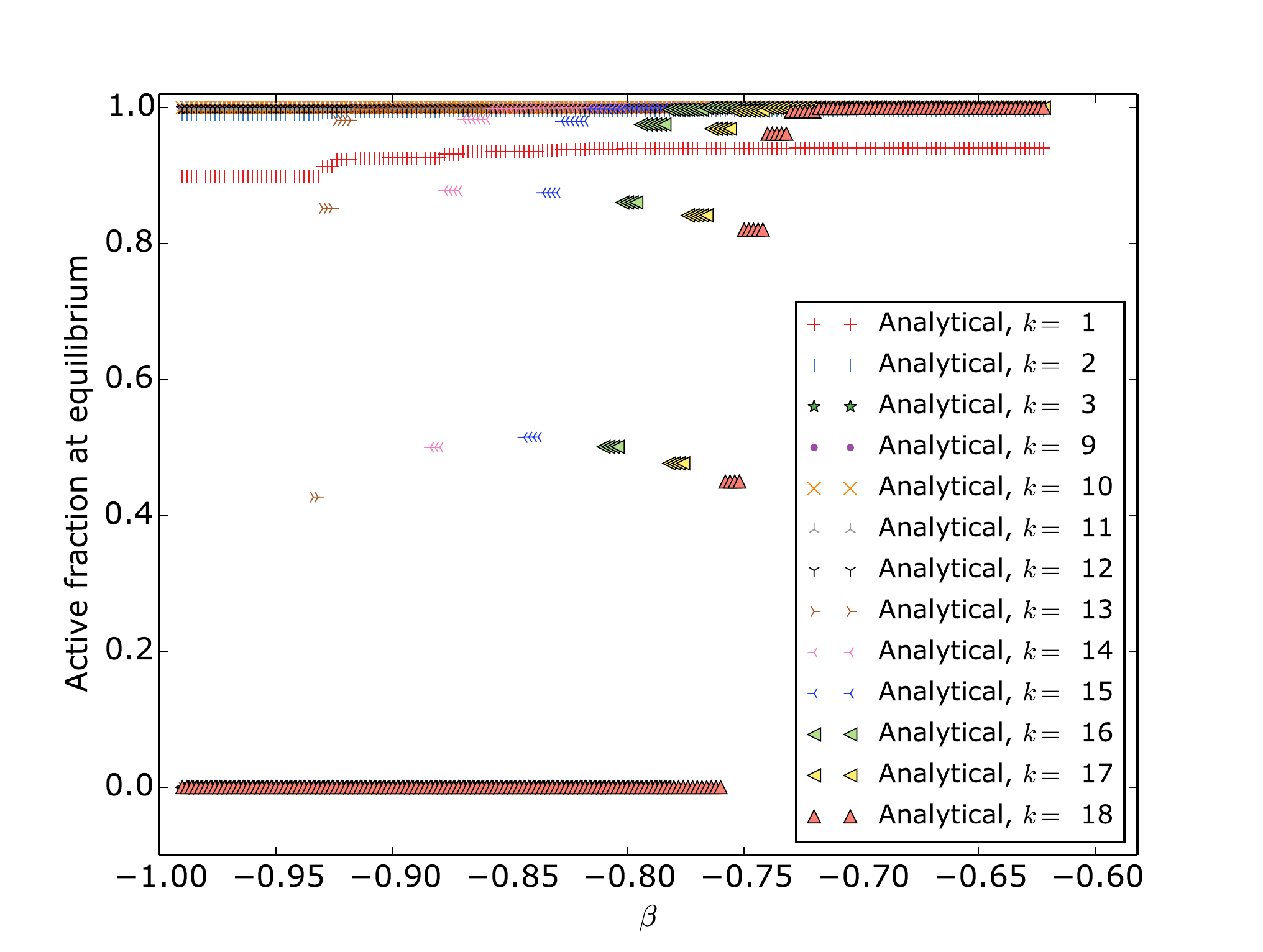}
\caption{Final active fraction of degree-$k$ nodes as a function of the synergy parameter $\beta$ for a meme that spreads on ER networks with mean degree $z=3$ and a response function with peer-pressure function \eqref{eq:power_func}. Using \eqref{eq:betacrit_power}, our analytical approximation \eqref{eq:rhok} again predicts abrupt jumps that match the calculations.
}
\label{fig:ER_z3_t_inf}
\end{figure}

\begin{figure}[tb]
\centering
\includegraphics[width=\linewidth]{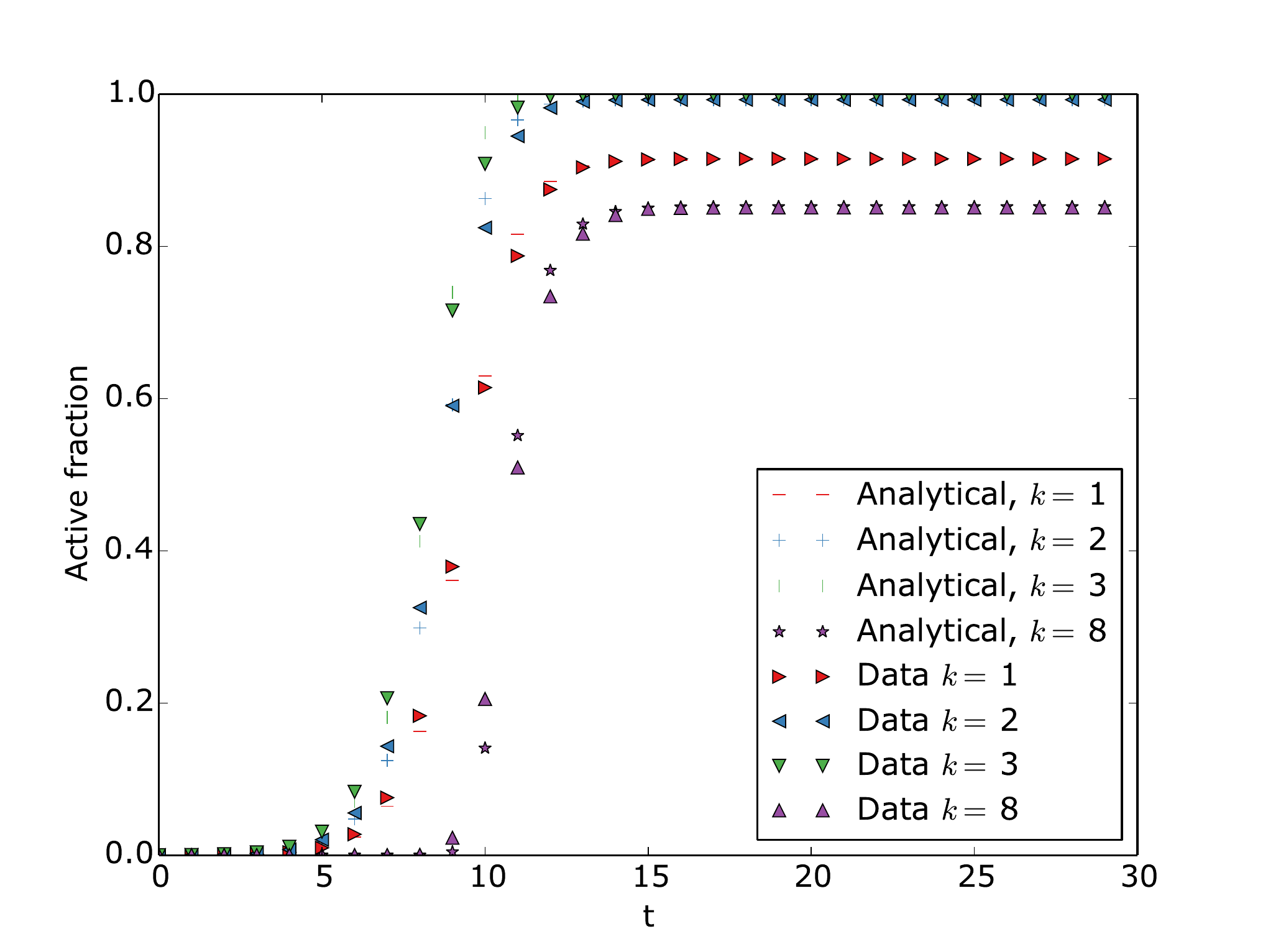}
\caption{Active fraction of nodes of degrees $1$, $2$, $3$, and $8$ nodes in ER networks as a function of time with synergy parameter $\beta=-0.93$, mean degree $z=3$, and homogeneous threshold $\phi = 1/7$. The memes spread using the synergistic response function with power peer-pressure function \eqref{eq:power_func}, and each data point is a mean over $31$ realizations of the spreading process. We observe that the analytical approximation \eqref{eq:rhok} of the temporal activation of nodes of degrees $1$, $2$, $3$, and $8$ match the simulated data well. This is also true for the other node degrees. We created a new random ER network was created for each realization. (In doing these simulations, we discarded $9$ realizations due to the choice of seed node; the contagion did not spread enough in those cases.)
}
\label{fig:ER_z3}
\end{figure}

We use our analytical approximation \eqref{eq:rhok} to find the expected equilibrium active fraction of nodes as a function of their degree and the synergy parameter $\beta$ for the response function with power peer-pressure function \eqref{eq:power_func}. 
We plot these quantities in Fig.~\ref{fig:ER_z3_t_inf}. In Fig.~\ref{fig:ER_z3}, we plot the time series of the fraction of active nodes when the symmetry parameter is $\beta = -0.93$, for which our model predicts different equilibrium active fractions for nodes with degrees $1$, $2$, $3$, and $8$.  We observe excellent agreement between our simulations and analytical approximation \eqref{eq:rhok} for these four node degrees.

\subsubsection{Mean Degree $z=8$}

\begin{figure}[tb]
\centering
\includegraphics[width=\linewidth]{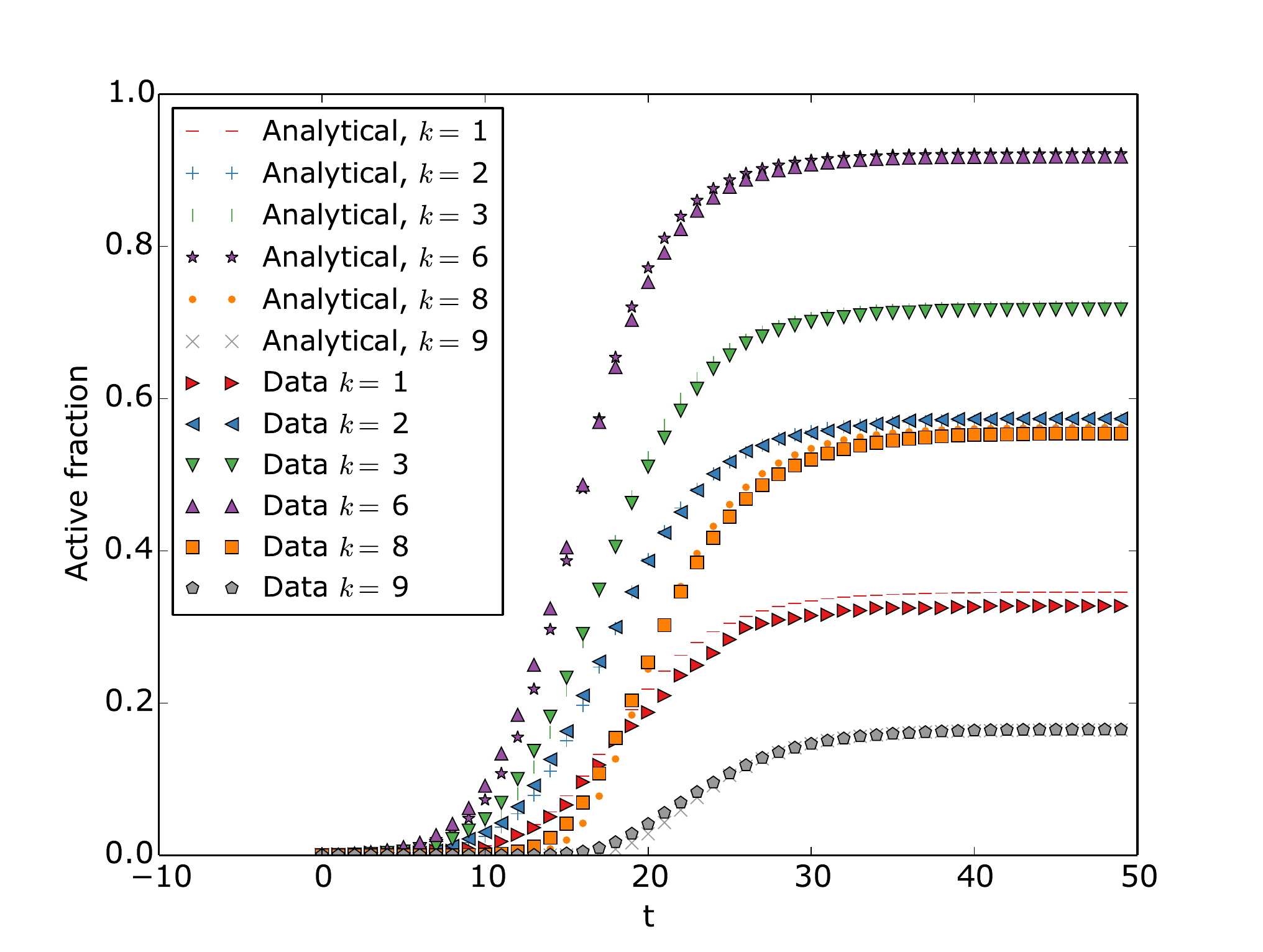}
\caption{Fraction of active degree-$k$ nodes as a function of time for ER networks with mean degree $z = 8$. We average the numerical results over $31$ realizations  of memes spreading over the networks using the synergistic response function with power peer-pressure function \eqref{eq:power_func} with $\beta = -0.835$. We observe a good match between our numerical results and our analytical approximation, although there is a slight discrepancy for nodes with $k=1$. (In doing these simulations, we discarded $9$ realizations due to the choice of seed node; the contagion did not spread enough in those cases.)
}
\label{fig:ER_z8}
\end{figure}

We now examine ER networks with mean degree $z=8$. We simulate the synergistic spreading processes with parameter $\beta = -0.835$ and a response function with power peer-pressure function \eqref{eq:power_func}.  
We choose this value of $\beta$ so that the final fraction of active nodes is different for nodes with different degrees. In Fig.~\ref{fig:ER_z8}, we show the fraction of active nodes as a function of time, and we again observe a good match between our computations and our analytical approximation \eqref{eq:rhok}.

\subsection{Synergy on Networks with Degree Distributions from Empirical Data}\label{sec:realistic}

We now simulate the spread of synergistic memes on two networks with degree distributions from empirical data.
In Section \ref{sec:CMP}, we consider random networks created using a configuration model (in particular, by matching stubs uniformly at random), in which we use a degree distribution from the degree sequence of the network of coauthorships in condensed-matter physics papers\cite{leskovec_graph_2007} that we examined in Section \ref{sec:initial}. This network has a mean degree of $z\approx 8$.  In Section \ref{sec:Facebook}, we simulate the spread of synergistic memes on networks created using a configuration model and a degree distribution from the degree sequence of the {\sc Northwestern25} network from the {\sc Facebook100} data set \cite{traud_social_2012,Traud2010}. 
This Facebook network has a mean degree of $z \approx 92$. For each realization, we create a new $10,000$-node network using a configuration model and degree sequences drawn from the associated degree distribution.


\subsubsection{Condensed-Matter Physics Collaboration Network}\label{sec:CMP}

\begin{figure}[tb]
\includegraphics[width=\linewidth]{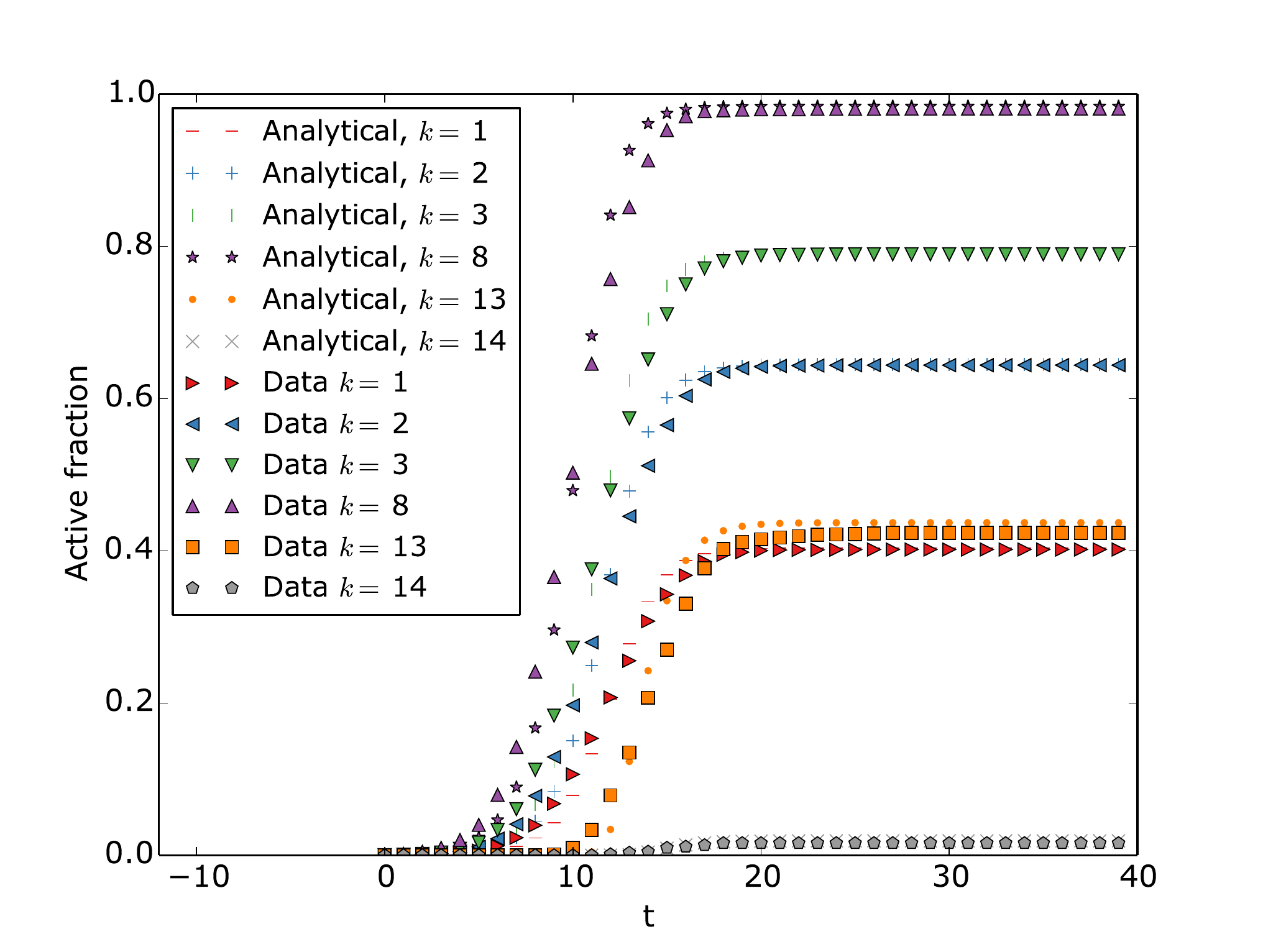}
\caption{Fraction of active nodes with degrees $1$, $2$, $3$, $8$, $13$, and $14$ as a function of time in configuration-model networks constructed from a degree distribution determined from the degree sequence of the condensed-matter theory collaboration network from \cite{leskovec_graph_2007}. We average the results over $9$ realizations of memes spreading using the synergistic response function with peer-pressure function \eqref{eq:power_func} and interfering synergy $\beta = -0.85$. Each node has a threshold of $\phi^*=1/10$. For each realization, we create a new configuration-model network. 
Our analytical approximation describes the results well.
(In doing these simulations, we discarded $1$ realization due to the choice of seed node; the contagion did not spread enough in those cases.)
}
\label{fig:CM}
\end{figure}

We draw the degree of each of the $10,000$ nodes from the degree distribution of the condensed-matter physics collaboration network \cite{leskovec_graph_2007}, and we create edges using a configuration model. 
In Fig.~\ref{fig:CM}, we plot the fraction of active nodes of degree $k$ as a function of time. We average over $9$ simulations (we discarded $1$ simulation because there was insufficient spreading from the seed node) of the spreading of a meme according to the power synergy model \eqref{eq:power_func} on condensed-matter collaboration networks. For each of these realizations, we create a new random network using a configuration model (as described above).

\begin{figure}[tb]
\includegraphics[width=\linewidth]{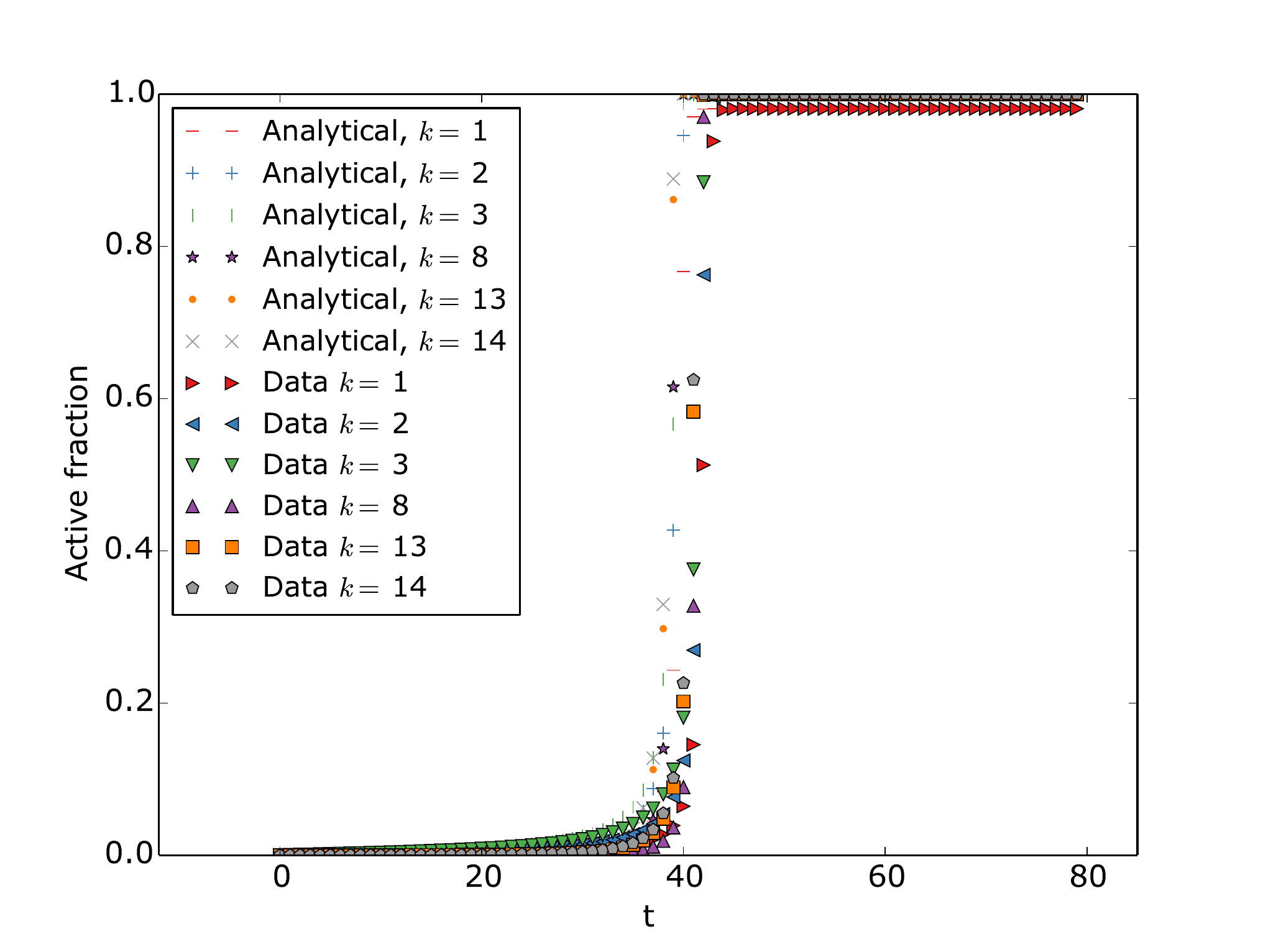}
\caption{Fraction of active nodes with degrees $1$, $2$, $3$, $8$, $13$, and $14$ as a function of time in configuration-model networks constructed from a degree distribution determined from the degree sequence of the condensed-matter theory collaboration network from \cite{leskovec_graph_2007}. We average the results over $10$ realizations of memes spreading using the synergistic response function with peer-pressure function \eqref{eq:power_func} and constructive synergy $\beta = 0.20$.
Each node has a threshold of $\phi^*=1/7$. Our analytical approximation predicts that the fraction of active nodes increases slightly earlier than what we observe in our numerical simulations, but the final fraction of infected nodes are visually indistinguishable in our analytical approximation and simulations. (In these simulations, we did not need to discard any realizations due to the choice of seed node.)
}
\label{fig:CM_constr}
\end{figure}

As in Section \ref{sec:initial}, we choose the threshold $\phi^* = 1/10$ for our simulations. We first consider interfering synergy with parameter $\beta = -0.85$, which makes it impossible to activate any node whose degree is $16$ or more. Our analytical approximation describes the simulated data well. In Fig.~\ref{fig:CM_constr}, we examine the effect of constructive synergy with the model \eqref{eq:power_func}. In this case, we use $\beta = 0.20$ and $\phi^* = 1/7$. For all node degrees that we checked, the final infected fraction is indistinguishable in the analytical prediction and the actual simulations. 
However, our analytical approximation predicts the infected fraction increases earlier than what occurs in our simulations.

\subsubsection{A Facebook Network}\label{sec:Facebook}

\begin{figure*}[tb]
\includegraphics[width=0.49\linewidth]{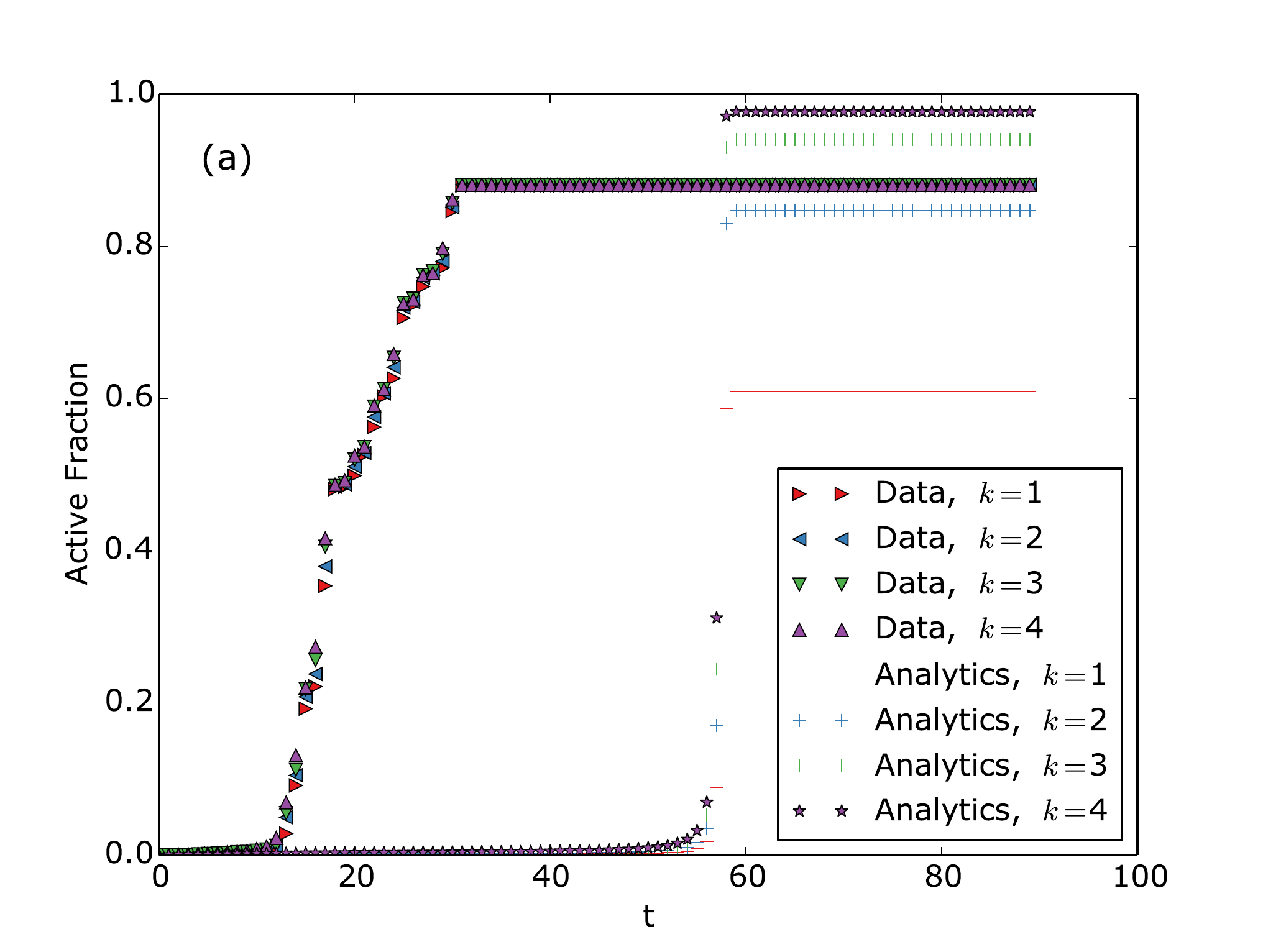}
\hfill
\includegraphics[width=0.49\linewidth]{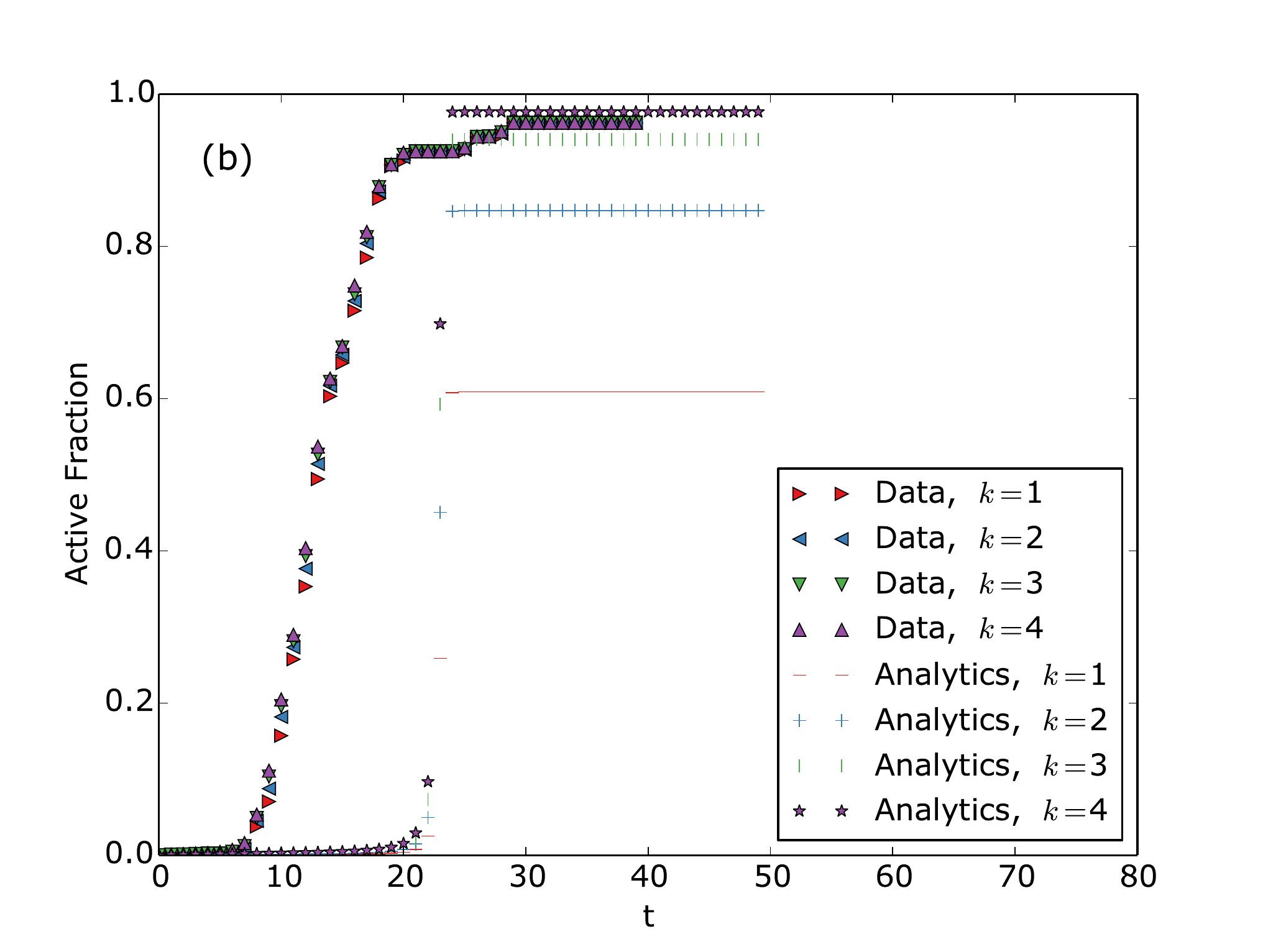}
\caption{Simulations of synergistic spreading processes on $10,000$-node networks with degree distribution determined from the {\sc Northwestern25} network from the {\sc Facebook100} data set \cite{traud_social_2012}. The nodes have a homogeneous threshold of $\phi^* = 1/33$. (a) We examine interfering synergy (with $\beta = -0.05$) and plot the fraction of active nodes with degrees $1$, $2$, $3$, and $4$ as a function of time. All nodes with degree $k\ge5$ exhibit behavior similar to those with the plotted degrees; the final fractions of active nodes are between $0.79$ and $0.88$. The time until the cascade occurs is very different between our analytical approximation \eqref{eq:rhok} and numerical simulations, and there are also discrepancies in the final fraction of active nodes between our analytics and numerics. We average our results over $51$ realizations. (In doing these simulations, we discarded $149$ realizations due to the choice of seed node; the contagion did not spread enough in those cases.)
 (b) We examine constructive synergy $\beta = 0.15$ and plot the fraction of active nodes with degrees $1$, $2$, $3$, and $4$ as a function of time. All nodes with degree $k\ge5$ eventually have fractions of active nodes that are larger than $0.92$. For this case as well, the time until the cascade occurs is very different in our analytical approximation \eqref{eq:rhok} and our numerical simulations, and there are also discrepancies in the final fraction of active nodes between our analytics and numerics. We average our results over $53$ realizations. (In doing these simulations, we discarded $147$ realizations due to the choice of seed node; the contagion did not spread enough in those cases.)
}
\label{fig:Facebook}
\end{figure*}

We simulate the spread of synergistic memes on the {\sc Northwestern25} network from the {\sc Facebook100} data set \cite{traud_social_2012}. The network has a mean degree of $z \approx 92$. The minimum degree is $d=1$, and the maximum degree is $d=2105$. We assign all nodes a degree from a degree distribution based on the degree sequence of the {\sc Northwestern25} network, and we create edges using a configuration model (in particular, by matching stubs uniformly at random, as we discussed previously). We suppose that each node has a threshold of $\phi^* = 1/33$. In Fig.~\ref{fig:Facebook}, we plot the fraction of active degree-$k$ nodes as a function of time. In panel (a), each data point is a mean over $51$ realizations of the spreading process on $10000$-node configuration-model networks with the degree distribution described above. (We discarded $149$ simulations because there was insufficient spreading from the seed node.) In panel (b), each data point is a mean over $53$ realizations. (We discarded $147$ simulations because there was insufficient spreading from the seed node.) As in our other simulations, each realization is a different draw of one of these configuration-model networks.

We show results for both interfering synergy (with $\beta = -0.05$) and constructive synergy (with $\beta = 0.15$). For this family of networks, our analytical approximation departs from our numerical simulations for both the final fractions of active nodes and the times at which the fractions of the active degree-$k$ nodes saturate. We also note that our analytical approximation suggests that interfering synergy slows down the spreading process much more than is actually the case in the simulations. 

Our analytical approximation assumes that we are considering dynamics on a locally tree-like network, although some processes often have ``ureasonable'' effectiveness even in many situations in which the hypotheses used to derive them fail to hold \cite{localtreeapprox_mason}. The authors of \cite{localtreeapprox_mason} discussed various reasons why a local-tree approximation may not provide a good description of the actual dynamics on a network (for a given dynamical system, such as a particular type of spreading process). For the {\sc Facebook100} networks, they found for various spreading processes (including the WTM) that simulations with a threshold distribution of $f(\phi) = \delta(\phi-\phi^*)$ yielded different dynamics in numerical simulations than in a tree-based theory. Reference \cite{localtreeapprox_mason} found with a Gaussian distribution of thresholds that simulations with a seed consisting of all nodes with $\phi<0$ yields results that are well-described by their local-tree approximation. In our case, however, altering the threshold distribution in this way does not yield agreement between our analytical approximation and simulated results.

Two properties that may provide some indication of how effectively tree-based theories for dynamical processes work on a network are the mean geodesic path length between nodes and a mean local clustering coefficient of the network. Although this is not something that is required mathematically (as one can construct counterexamples, so as a star graph), it reasonable to expect an ``typical'' tree-like network (in particular, consider an ensemble of networks drawn uniformly at random from the set of all trees with a given number of nodes) to have larger mean geodesic path lengths than ``typical'' networks of the same size that are not tree-like (e.g., an ensemble of configuration-model networks). One also expects a tree-like network to have a smaller mean local clustering coefficient than a network with the same number of nodes that is not tree-like. Averaging the mean geodesic path length between nodes in a set of $10$ randomizations (as described above) of the {\sc Northwestern25} network yields $2.510\pm 0.007$, which is much lower than any other random network in this study (see Table \ref{tab:averaged_path_length}). Averaging the local clustering coefficient for the same $10$ networks yields $0.02828\pm0.00109$, which is much higher than any other random network in this study. This suggests that the randomized {\sc Northwestern25} networks are less tree-like than the other random networks that we examine. Additionally, the mean local clustering coefficient and the mean geodesic path length in the original {\sc Northwestern25} and condensed-matter collaboration networks are larger than those of the randomized networks that we constructed from those networks. Unsurprisingly, randomization considerably decreases the value of the mean local clustering coefficients (especially for the condensed-matter collaboration network).

\begin{table*}[tb]
\begin{tabular}{|l|c|c|}
\hline
\textbf{Network} & \textbf{Mean geodesic path length}
& \textbf{Mean local clustering coefficient}\\
\hline
$3$-regular & $6.359 \pm 0.001$ &$0.00033 \pm 0.00011$\\
\ER {}, $z = 3$  & $8.366 \pm 0.043$& $0.00020 \pm 0.00014$
\\
\ER {}, $z = 8$  & $4.664 \pm 0.009$& $0.00079 \pm 0.00008$
\\
Cond-mat collab. (original) & $ 5.352$
& $0.64173$
\\
Cond-mat collab. (random) & $ 4.091 \pm 0.024$&$0.00471 \pm 0.00049
$ \\
{\sc Northwestern25} (original) & $2.723$
& $0.23828$
\\
{\sc Northwestern25} (random) & $2.509 \pm 0.007$& $0.02828 \pm 0.00109$\\
\hline
\end{tabular}
\caption{Mean geodesic path length between nodes and mean local clustering coefficient in the four different random-network families that we examine. We construct the 3-regular random graphs using the configuration model (with stubs connected uniformly at random), and we also construct configuration-model networks using degree sequences (with associated degree distributions) from the condensed-matter collaboration network and {\sc Northwestern25} Facebook network.
In each case, we average our results over 10 networks, and we indicate the mean values and the standard deviation of mean in each case. We also list the values for the original {\sc Northwestern25} and condensed-matter collaboration networks. Observe that the mean geodesic path length between nodes is much smaller in the {\sc Northwestern25} networks than in the other networks. Among the random networks, the mean local clustering coefficient is also by far the largest in the {\sc Northwestern25} network, although the condensed-matter collaboration network also has a much larger mean local clustering coefficient compared to the ER networks and the $3$-regular networks. The original empirical networks have values for both the mean geodesic path length and mean local clustering coefficients that are significantly larger than the values in the random networks with degrees drawn from the same degree distributions. 
}
\label{tab:averaged_path_length}
\end{table*}

\section{Conclusions}\label{conc}

It is important to study when diseases, information, a meme, or something else (e.g., misinformation or ``alternative facts'') spreads to a large number of nodes in a network \cite{fowlerreview,porter2016}. Prior studies have suggested that some organisms and tumors spread via synergistic effects \cite{ben1994generic, liotta2001microenvironment} and that synergistic effects can also be important for the spread of information on networks \cite{PhysRevE.88.012818}, the spread of behavior in online social networks \cite{Centola1194}, the transmission of pathogens \cite{ludlam2012applications}, and the spread of opportunities among vineyards on wine routes\cite{brunori2000synergy}. 

In the present paper, we developed a threshold model with synergistic spreading; and we investigated both analytically and computationally the fraction of nodes, resolved by degree and as a function of a synergy parameter, that are activated for empirical networks and several families of random graphs. We illustrated that the synergistic models \eqref{eq:betacrit_multi} and \eqref{eq:betacrit_power} lead to critical synergy levels at which non-vulnerable nodes with a certain degree $k$ can be activated by $k$ active neighbors for all synergy parameter values of at least this level.

We used a local-tree approximation to approximate the fraction of active degree-$k$ nodes as a function of time. We illustrated that our analytical approximation \eqref{eq:rhok} matches well with numerical simulations for synergistic memes that spread on $3$-regular random networks, \ER {} networks, and configuration-model networks constructed from a condensed-matter physics collaboration data set. However, our analytical approximation does not do well for configuration-model networks that we construct using a degree distribution of the {\sc Northwestern25} data set. We pointed out that the random networks constructed from the {\sc Northwestern25} network differ from the other networks that we examined in that, on average, they have a much larger mean local clustering coefficient and a much shorter mean geodesic path length. In all cases, we observed that constructive synergy speeds up the spreading process and that interfering synergy slows down the spreading process. 

The influence of synergistic effects on spreading processes in networks is a promising area of study.  It is an important component of modeling the spread of information (and misinformation) on social networks \cite{PhysRevE.88.012818} and the behavior of certain biological organisms and social processes in which a willingness to adopt either saturates or increases with the number of individuals who are trying to influence other individuals in a network. It has interesting effects on spreading behavior in various types of networks, such as lattices versus other networks\cite{PhysRevE.88.012818} and in modular networks\cite{PhysRevE.89.052811}, and it can affect whether it is possible or impossible for certain nodes to adopt a certain meme or behavior. 

In the future, it will be interesting to consider synergistic spreading processes on other types of networks, such as multilayer networks \cite{domen2016,salehi2015spreading,Kivela2014} and temporal networks \cite{Holme2012}.


\section*{Acknowledgements}

This work was carried out at the Mathematical Institute at University of Oxford. We thank James Fowler, James Gleeson, Matthew Jackson, Mikko Kivel\"a, and James Moody for helpful discussions.  JSJ also thanks the Mathematical Institute, University of Oxford for their hospitality and Mogens H\o gh Jensen (Niels Bohr Institute, University of Copenhagen) for making this project possible.






\end{document}